\documentclass[english,compress,square,super]{elsarticle}

\usepackage{lineno}
\modulolinenumbers[1]
\usepackage{geometry}
\usepackage[colorlinks=true]{hyperref}
\journal{arXiv.}

\usepackage{amsmath}
\usepackage{xcolor}
\newcommand\tb[1]{\boldsymbol{#1}}
\newcommand\td{\mathrm{d}}

\usepackage{mathrsfs}

\newcommand\ddfrac[2]{{\displaystyle\frac{\displaystyle #1}{\displaystyle #2}}}
\usepackage{stmaryrd}
\usepackage{enumitem}
\usepackage{caption}
\usepackage{subcaption}
\usepackage{hyperref}
\usepackage{cleveref}
\Crefname{figure}{\text{Fig.}}{\text{Figs.}}
\usepackage{algorithm}  
\usepackage{algpseudocode}  
\usepackage{amssymb}
\usepackage{upgreek}
\usepackage{pgfplots}
\pgfplotsset{compat=1.16}
\usepackage{svg}
\usepackage{multirow}
\usepackage{booktabs}
\usepackage{mathtools}
\usepackage{adjustbox}
\usepackage[para,online,flushleft]{threeparttable}

\makeatletter
\usepackage{xspace}
\algnewcommand{\algorithmicgoto}{\textbf{go to}}%
\algnewcommand{\Goto}{\algorithmicgoto\xspace}%
\DeclareRobustCommand\bigop[1]{%
  \mathop{\vphantom{\sum}\mathpalette\bigop@{#1}}\slimits@
}
\newcommand{\bigop@}[2]{%
  \vcenter{%
    \sbox\z@{$#1\sum$}%
    \hbox{\resizebox{\ifx#1\displaystyle.9\fi\dimexpr\ht\z@+\dp\z@}{!}{$\m@th#2$}}%
  }%
}
\makeatother

\newcommand{\revision}[1]{\textcolor{black}{#1}}

\usepackage{amsthm}
\newtheorem{remark}{Remark}

\begin{document}
\sloppy

\begin{frontmatter}

\title{Absorbing boundary conditions in material point method adopting perfectly matched layer theory}

\author[mysecondaryaddress]{Jun Kurima}
\ead{kurima@iis.u-tokyo.ac.jp}

\author[mymainaddress,mysecondaddress]{Bodhinanda Chandra\corref{cor1}}
\ead{bchandra@berkeley.edu}
\cortext[cor1]{Corresponding author}

\author[mymainaddress]{Kenichi Soga}
\ead{soga@berkeley.edu}

\address[mysecondaryaddress]{Institute of Industrial Science, The University of Tokyo, Japan}

\address[mymainaddress]{Department of Civil and Environmental Engineering, University of California, Berkeley, CA, 94720, USA}

\address[mysecondaddress]{Department of Mechanical Engineering, University of California, Berkeley, CA, 94720, USA}

\begin{abstract}
This study focuses on solving the numerical challenges of imposing absorbing boundary conditions for dynamic simulations in the material point method (MPM). To attenuate elastic waves leaving the computational domain, the current work integrates the Perfectly Matched Layer (PML) theory into the implicit MPM framework. The proposed approach introduces absorbing particles surrounding the computational domain that efficiently absorb outgoing waves and reduce reflections, allowing for accurate modeling of wave propagation and its further impact on geotechnical slope stability analysis. The study also includes several benchmark tests to validate the effectiveness of the proposed method, such as several types of impulse loading and symmetric and asymmetric base shaking. The conducted numerical tests also demonstrate the ability to handle large deformation problems, including the failure of elasto-plastic soils under gravity and dynamic excitations. The findings extend the capability of MPM in simulating continuous analysis of earthquake-induced landslides, from shaking to failure.
\end{abstract}

\begin{keyword}
Material point method\sep Perfectly matched layer \sep Absorbing boundary \sep Dynamic analysis \sep Slope stability \sep Earthquake-induced landslides
\end{keyword}

\end{frontmatter}


\section{Introduction}
Seismic activities have long posed significant challenges in geotechnical engineering, particularly due to their potential to trigger landslides. Notably, the 2008 Sichuan earthquake in China and the 2015 Gorkha earthquake in Nepal have demonstrated the consequences of seismic-induced landslides, leading to substantial human, social, and economic losses \cite{wenchuan1, wencuan2, gorkha1,gorkha2}. These events underscore the need for advanced analysis methods capable of accurately predicting the impacts of seismic activities on slope stability. 

Traditionally, the assessment of earthquake-induced landslides has relied on classical analysis methods \cite{jibson}, including the Limit Equilibrium Method (LEM) \cite{lem1,lem2,lem3}, Newmark's sliding block analysis \cite{Newmark,newmark2,newmark3,newmark4}, and the Finite Element Method (FEM) \cite{fem1,fem2}. While these methodologies have significantly contributed to our understanding and assessment capabilities, they each have inherent limitations. LEM is a static approach used to assess the stability of slopes and ground masses. It cannot directly handle dynamic responses, such as those caused by earthquakes. LEM assumes potential failure mechanisms along discontinuous surfaces (slip surfaces) and analyses their equilibrium conditions. As a result, it does not capture the detailed processes of continuous deformation and failure within the soil mass \cite{jibson}. Meanwhile, Newmark's method approximates the slope material as rigid bodies, hence, neglecting internal deformation, stress distribution, as well as its elasto-plastic behavior. This can lead to inaccurate predictions, especially in scenarios where material softens upon shearing \cite{deng,jafarian,biondi}. On the other hand, it is widely recognized that FEM faces numerical challenges in large-deformation regimes, such as mesh entanglements \cite{strain1,strain2,strain3}, which limit its effectiveness in dealing with the post-failure mechanism observed in many geotechnical engineering scenarios. Several mitigation strategies such as the Arbitrary Lagrangian-Eulerian (ALE) \cite{ALE1,ALE2}, the Coupled Euler-Lagrangian (CEL) \cite{cel1,cel2}, and the Smoothed Finite Element Method (SFEM) \cite{sfem1,sfem2,sfem3} have been proposed in the past to alleviate the mesh entanglement issues. However, other issues emerged within these methods including mass conservation, numerical diffusion, and proper tracking of the evolution of history-dependent variables \cite{soga2016}.

The mesh-free methods provide other alternatives to handle the limitations of traditional methods, such as the Discontinuous Deformation Analysis (DDA) \cite{dda1,dda2}, for the analysis of rock structures, and the Smoothed Particle Hydrodynamics (SPH) for the analysis of continuum materials. These approaches offer flexibility in modeling large deformations and complex material behaviors without any mesh handling. In the past few years, many research works have been developed to handle issues related to mesh-free methods to impose boundary conditions (BCs). For geomechanical analysis, the boundary conditions proposed by \citet{BUI2008} are widely considered due to their effectiveness in handling inhomogeneous displacement boundaries. However, challenges persist in addressing boundary conditions for dynamic settings. While some strategies for this aspect have been proposed for fluid wave applications in the past (e.g.~\cite{negi,molteni}), boundary formulations for seismic analysis are still limited. Recent efforts, such as introducing dashpot elements to dissipate stress waves at the boundary, offer some progress \cite{hoang2024development}.

The material point method (MPM) has been proposed by \citet{sulsky1994} and developed further as a method capable of handling large-deformation solid mechanics problems. The method utilizes a hybrid approach that combines the advantages of both mesh-based and mesh-free methods. Over the past two decades, the MPM has garnered significant attention to model many challenges in geomechanics \cite{soga2016,bandara2015coupling, larese2019implicit, kularathna2021semi, baumgarten2021coupled, liang2022shear, yu2024semi}, especially in scenarios like landslides and other mass movement events. One of the major difficulties encountered in MPM is simulating mass movement scenarios induced by earthquakes or other types of ground vibrations. This is primarily because appropriate modeling of far-field boundary conditions is necessary to ensure that waves traveling outward from the simulation domain do not get refracted back, and thus, should be properly damped or absorbed. These types of boundary conditions prevent the energy generated by ground vibrations from remaining within the simulation domain, thereby avoiding unrealistic accumulation that can lead to excessive deformation \cite{higo}.

The implementation of dynamic boundary conditions to prevent wave reflections within the simulation domain is paramount. Typically, artificial boundaries based on rheological combinations of spring and dashpot (e.g.~\citet{lysmer}) are introduced primarily to mitigate reflections by imposing viscous damping force on the boundary. The use of these boundary conditions in Lagrangian FEM and other mesh-based methods is rather straightforward, as the material boundaries always conform with the Lagrangian mesh and domain boundaries. However, a well-known challenge in MPM is that the material boundaries are often incompatible with the domain and mesh boundaries \cite{NAnda2021}, hence, the application of dynamic boundaries is not trivial, particularly when large shaking is involved. Many MPM studies in the past have used some types of the aforementioned viscous boundaries, e.g.~\cite{kafaji2013formulation, jassim2013two, kohler,gangwang1,gangwang2,silentmpm,dampingtwophase}, even though it has been noted that viscous boundaries alone will not be able to absorb reflected waves sufficiently \cite{zhaopml,zhangpml}. Periodic boundaries are also often employed in earthquake analysis, which has been employed for seismic analysis in MPM by \citet{alba}. However, periodic boundaries present numerical limitations, including the restriction of using spatially invariant input waves to the bottom boundary of the model, as well as their inapplicability for geometrically asymmetric models, such as conditions with varying elevations or subsurface profiles. 

Due to the aforementioned difficulties and limitations, many recent and prior studies have attempted to integrate dynamic earthquake analysis using FEM or other mesh-based methods with transient large-deformation analysis using MPM \cite{talbot2024modeling, sordo4783551sequential,fem-mpm}. However, this approach raises two fundamental questions: (i) \emph{do landslides occur after earthquake shaking has ceased or during the shaking}? and (ii) \emph{when is the appropriate condition to transition from the initial dynamic analysis to MPM, ensuring the accurate mapping of necessary quantities}? Naturally, landslides are triggered during the earthquake, making the concept of subsequent analysis generally a simplification that divides continuous landslide-triggering phenomena into two phases. As for the second question, it is challenging to establish a general rule. Some may argue that the transition should occur when the initial method fails to converge, e.g.~due to mesh entanglement. However, others may contend that the numerical solution near the mesh-entanglement state is erroneous, suggesting that the transition should take place earlier \cite{sordo4783551sequential}. At the same time, faster transfer time may remove dynamic effects towards material deformation, which is only modeled in the first analysis. A clear methodology defining the most optimal time to transition from the initial analysis to the subsequent MPM analysis has yet to be established.

Another alternative concept available in the literature is the absorbing boundary condition, which simulates waves reaching the boundary and then progressing into an external \textit{semi-infinite space}. There are two main classes of this method \cite{hori}: (i) one that modifies the boundary differential equations to eliminate reflected waves at artificial boundaries \cite{clayton} and (ii) another that places an artificial body next to the boundary to ensure rapid wave attenuation. Belonging to the second category, the Perfectly Matched Layers (PML) have been developed as an efficient method for absorbing outgoing waves \cite{pled2022}. Initially introduced by \citet{berenger} for electromagnetic wave simulations, PML has since been adapted and extensively studied for various wave equations, including those governing elastic wave behaviors, making it a preferred tool in seismic wave propagation simulations \cite{kumar,ye,wang}. Theoretically, PML absorbs waves at all incident angles and frequencies without generating reflective waves using coordinate stretching in the frequency domain. Therefore, facilitating the use of fixed boundaries. A recent review of PML development has been nicely summarized by \citet{pled2022}.

\begin{figure}[h!]
    \centering
    \includegraphics[width=0.9\linewidth]{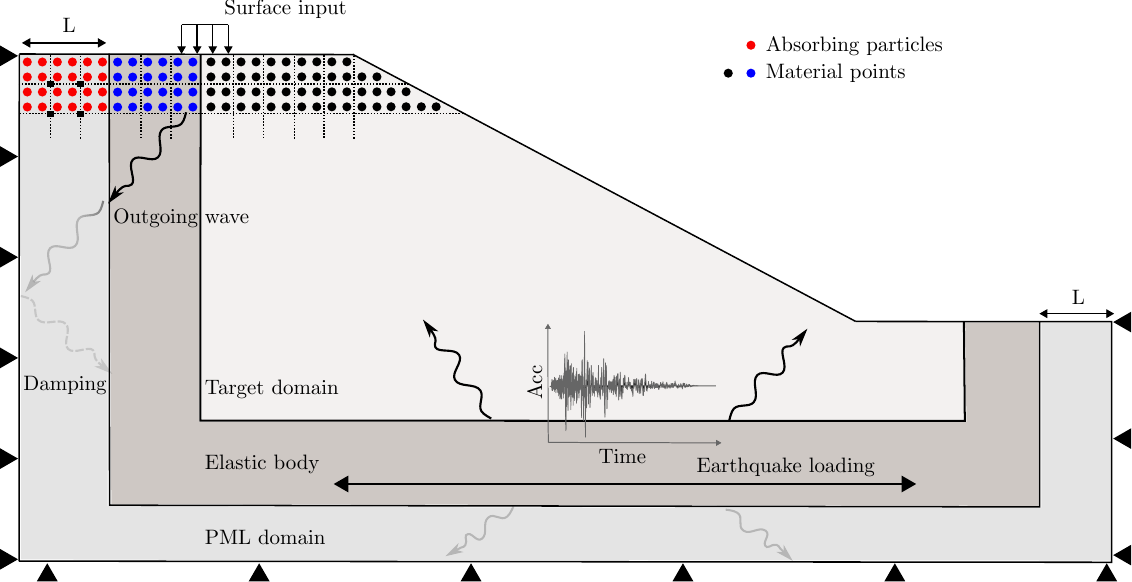}
    \caption{Illustration of MPM computational model for dynamic analysis using PML.}
    \label{fig:scheme}
\end{figure}

Our study aims to advance the integration of PML into the implicit displacement-based MPM framework by introducing a set of \textit{absorbing particles}. The proposed formulation was built upon the formulation proposed by \citet{chen2022} as their formulation can be easily implemented within the time domain Galerkin framework. As depicted schematically in Fig.~\ref{fig:scheme}, reflected waves are attenuated within the PML domain, and fixed displacement boundary conditions can be enforced at the domain boundaries. Additionally, the computational model can be asymmetric and the various types of input loads can be applied. The proposed method also allows seamless integration of dynamic analysis with elasto-plastic analysis of materials under large deformation conditions. The developed framework is then validated against various loading conditions, including impulse loading, symmetric base shaking, and asymmetric base shaking, to demonstrate its effectiveness and potential. 

In this paper, we first present the governing equations and implementation details that introduce PML into the MPM in \Cref{intro_pml}. Then, \Cref{sec:num_examples} presents the conducted numerical tests to assess the stability and validity of the proposed method. In \Cref{elastic_2d}, a numerical analysis of an elastic body with pulse-like loading on the ground surface is analyzed, whereas in  \Cref{elasticplastic_slope}, an elasto-plastic symmetrical embankment is modeled with input ground motions of various amplitudes. Furthermore, in \Cref{slope_failure}, the time evolution and post-failure deformation of an asymmetric slope are investigated and compared with existing numerical results performed by other boundary treatments. Lastly, \Cref{sec:conclusion} concludes the study and provides suggestions for future works.

\section{Perfectly-matched-layer (PML) boundary in implicit MPM}\label{intro_pml}

In this section, we introduce the PML element formulation proposed by \citet{chen2022} and its adoption to implicit MPM. Before delving into the mathematical formulations, let us first establish some useful notations. Here, we introduce three types of subscripts or superscripts: $\square_p$ for material point or particle variables, $\square_I$ or other capital alphabet subscripts for nodal variables, and $\square^t$ for time indices, where $t=n$ and $t=n+1$ denote current and future variables, respectively. We consider a few mathematical operators: $\dot{\square}$ and $\ddot{\square}$ indicate the first and second-order material time derivatives, $\square\cdot\square$ and $\square:\square$ indicate the single and double contractions of tensor indices, and $\square\otimes\square$ indicates the tensorial dyadic operator. 
The subscript $j$ or $k=\{1,2,3\}$ are often used to represent the directional index associated with $x$, $y$, and $z$-coordinates. This paper does not employ Einstein's summation convention in expressions with repetitive indices.

\subsection{Time-domain PML formulation}
The three-dimensional elastic wave equation can be expressed as the following equation for each direction $j$: 
\begin{eqnarray}\label{eq:wave_equation3d}
\begin{aligned}
\rho  \ddot{u}_j&=\frac{\partial}{\partial x_j}\left[(\lambda+2 G) \frac{\partial u_j}{\partial x_j}\right]+ \sum_{\substack{k=1 \\ k \neq j}}^3 \Biggl\{ \frac{\partial}{\partial x_j}\left[\lambda \frac{\partial u_k}{\partial x_k}\right]+\frac{\partial}{\partial x_k}\left[G\left( \frac{\partial u_k}{\partial x_j}+  \frac{\partial u_j}{\partial x_k}\right)\right]\Biggl\}\,,
\end{aligned}
\end{eqnarray}
where $\rho$ is the mass density and $\lambda$ and $G$ are the Lame's constants. In the above notation, $u_j$ is the displacement vector component in the $j$-th direction. Performing Fourier transform to Eq.~\eqref{eq:wave_equation3d} yields the following expressions:
\begin{eqnarray}\label{eq:wave equation3d fourier}
\begin{aligned}
\rho (i\omega)^2 \hat{u}_j&=\frac{\partial}{\partial x_j}\left[(\lambda+2 G) \frac{\partial \hat{u}_j}{\partial x_j}\right]+ \sum_{\substack{k=1 \\ k \neq j}}^3 \Biggl\{ \frac{\partial}{\partial x_j}\left[\lambda \frac{\partial \hat{u}_k}{\partial x_k}\right]+\frac{\partial}{\partial x_k}\left[G\left( \frac{\partial \hat{u}_k}{\partial x_j}+  \frac{\partial \hat{u}_j}{\partial x_k}\right)\right]\Biggl\} \,.
\end{aligned}
\end{eqnarray}
Here, $i$ and $\omega$ denote the 
imaginary number and the angular frequency, respectively, whereas $\hat{u}_j$ denotes the Fourier-transformed displacement vector component.   

In PML theory, the wave equation is transformed using the concept of complex coordinate stretching. This transformation is described using a directional-dependent damping function $C_j$. The complex coordinate stretching is expressed using $C_j$ as follows:
\begin{eqnarray}
\label{eq:damping functions}
\begin{aligned}
\widetilde{x}_j &= x_j - \frac{i}{\omega} \int_{x_{j_0}}^{x_j} C_j \, \td x_j \quad (\text{in the } j\text{-th} \text{ direction})\,.
\end{aligned}
\end{eqnarray}
We can also obtain the partial derivative of Eq.~\eqref{eq:damping functions} as:
\begin{eqnarray}
\label{eq:partial dumping functions}
\begin{aligned}
\frac{\partial }{\partial \widetilde{x}_j} &= \frac{i\omega}{i \omega +C_j} \frac{\partial }{\partial x_j} \,,
\end{aligned}
\end{eqnarray}
and by replacing all the spatial terms and derivatives in Eq.~\eqref{eq:wave equation3d fourier} with Eqs.~\eqref{eq:damping functions}-\eqref{eq:partial dumping functions}, and performing certain mathematical manipulations, we can obtain the following expression written in the frequency domain:
\begin{eqnarray}
\label{eq:fourier governing equations}
\begin{aligned}
\rho \hat{u}_j &= \frac{1}{(i \omega+C_j)^2} \frac{\partial}{\partial x_j}\left[(\lambda+2G) \frac{\partial \hat{u}_j}{\partial x_j}\right] +\sum_{\substack{k=1 \\ k \neq j}}^3 \Biggl\{ \frac{1}{(i \omega+C_j)(i \omega+C_k)} \frac{\partial}{\partial x_j}\left[\lambda \frac{\partial \hat{u}_k}{\partial x_k}\right]\\
&\qquad +\frac{1}{(i \omega+C_j)(i \omega+C_k)} \frac{\partial}{\partial x_k}\left[G \frac{\partial \hat{u}_k}{\partial x_j}\right]+\frac{1}{(i \omega+C_k)^2} \frac{\partial}{\partial x_k}\left[G \frac{\partial \hat{u}_j}{\partial x_k}\right]\Biggl\}\,.
\\
\end{aligned}
\end{eqnarray}

To simplify Eq.~\eqref{eq:fourier governing equations}, following \citet{chen2022}, it is assumed that
\begin{eqnarray}
\label{eq:assumpitions for dumping functions}
C_j=i \omega C_j^{\prime}\,,
\end{eqnarray}
and by substituting Eq.~\eqref{eq:assumpitions for dumping functions} to Eq.~\eqref{eq:fourier governing equations}, we can rewrite it to:
\begin{eqnarray}\label{eq:dumping before inverse}
\begin{aligned}
\rho\left(i\omega+ i\omega C_j^{\prime}\right)^2\hat{u}_j&=\frac{\partial}{\partial x_j}\left[(\lambda+2 G) \frac{\partial \hat{u}_j}{\partial x_j}\right] \\
&+ \sum_{\substack{k=1 \\ k \neq j}}^3 \Biggl\{\left(1+C_j^{\prime}\right) \left(\frac{\partial}{\partial x_j}\left[\lambda \frac{\partial \hat{u}_k}{\partial x_k}\right]+\frac{\partial}{\partial x_k}\left[G \frac{\partial \hat{u}_k}{\partial x_j}\right]\right)+\left(1+C_j^{\prime}\right)^2 \frac{\partial}{\partial x_k}\left[G \frac{\partial \hat{u}_j}{\partial x_k}\right]\Biggl\}\,.
\end{aligned}
\end{eqnarray}
Notice that, in the above expression, we consider that $C'_k=0$ for a fixed direction $j$.

Furthermore, we perform an inverse Fourier transform of Eq.~\eqref{eq:dumping before inverse} to obtain the following modified governing equation in the time domain:
\begin{eqnarray}\label{eq:dumping after inverse}
\begin{aligned}
\rho\left(1+C_j^{\prime}\right)^2  \ddot{u}_j&=\frac{\partial}{\partial x_j}\left[(\lambda+2 G) \frac{\partial u_j}{\partial x_j}\right]\\
&+ \sum_{\substack{k=1 \\ k \neq j}}^3 \Biggl\{\left(1+C_j^{\prime}\right) \left(\frac{\partial}{\partial x_j}\left[\lambda \frac{\partial u_k}{\partial x_k}\right]+\frac{\partial}{\partial x_k}\left[G \frac{\partial u_k}{\partial x_j}\right]\right)+\left(1+C_j^{\prime}\right)^2 \frac{\partial}{\partial x_k}\left[G \frac{\partial u_j}{\partial x_k}\right]\Biggl\}\,.
\end{aligned}
\end{eqnarray}
It is worth emphasizing that the original elastic wave equation can be recovered by setting the damping function above to zero, i.e.~by setting $C'_j=0$, Eq.~\eqref{eq:dumping after inverse} will in fact be Eq.~\eqref{eq:wave_equation3d}. It is also important to note that the equation above is direction-dependent because of the direction variability of the damping function.

The damping function utilized in the current work is defined as follows \cite{basu2009,bindel}:
\begin{eqnarray}
 {C}_j^{\prime}=\alpha_j\left(\frac{\left\vert x_j-x_{j_b}\right\vert}{L_j}\right)^\beta =\alpha_j\left(\frac{d_j}{L_j}\right)^\beta\,,  
 \label{eq:damping_func}
\end{eqnarray}
where $\alpha_j$ denotes the maximum values of ${C}_j^{\prime}$ at the outermost boundary of the absorbing region (also referred to as PML domain in Fig.~\ref{fig:scheme}) with a predefined thickness, $L_j$. The coordinates $x_{jb}$ denote the location of the inner boundary of this absorbing region, while the superscript $\beta$ indicates the polynomial power. In Eq.~\eqref{eq:damping_func}, we also define a positive distance function $d_j=\left\vert x_j-x_{j_b}\right\vert$ with range of $0\leq d_j\leq L_j$, such that the damping function is zero at the inner boundary and the non-PML domain, and $d_j=\alpha_j$ at the outer boundary. In the current work, $d_j$ is predefined at the beginning of the simulation and its value is fixed in the entire simulations.

\subsection{Weak formulation}
By applying the standard Galerkin approximation to Eq.~\eqref{eq:dumping after inverse}, the weak form of momentum balance can be written in a residual form as:
\begin{eqnarray}\label{eq:weak form first}
\begin{aligned}
{R}_j(\tb u,\delta u_j) =&\int_\Omega \rho\left(1+C_j^{\prime}\right)^2  \ddot{u}_j \delta u_j \td V + \int_\Omega \left(\frac{\partial}{\partial x_j}\left[(\lambda+2 G) \frac{\partial u_j}{\partial x_j}\right]\right) \delta u_j \td V \\
&+\int_\Omega\sum_{\substack{k=1 \\ k \neq j}}^3 \Biggl\{\left(1+C_j^{\prime}\right) \left(\frac{\partial}{\partial x_j}\left[\lambda \frac{\partial u_k}{\partial x_k}\right]+\frac{\partial}{\partial x_k}\left[G \frac{\partial u_k}{\partial x_j}\right]\right)+\left(1+C_j^{\prime}\right)^2 \frac{\partial}{\partial x_k}\left[G \frac{\partial u_j}{\partial x_k}\right]\Biggl\} \delta u_j \td V =0\,,
\end{aligned}
\end{eqnarray}
where $\delta u_j$ denotes the arbitrary displacement test function  in the direction $j$, while $\Omega$ denotes the computational domain. Omitting the traction term is for simplicity, Eq.~\eqref{eq:weak form first} can be rewritten as:
\begin{eqnarray}\label{eq:weak form sec}
     \begin{aligned}
     {R}_j(\tb u,\delta u_j)=& \int_\Omega \rho\left(1+C_j^{\prime}\right)^2 \ddot{u}_j \delta u_j \td V +\int_\Omega (\lambda+2 G) \frac{\partial u_j}{\partial x_j}\frac{\partial \delta u_j}{\partial x_j} \td V\\
&\qquad +\int_\Omega\sum_{\substack{k=1 \\ k \neq j}}^3\left(1+C_j^{\prime}\right) \left(\lambda \frac{\partial u_k}{\partial x_k}\frac{\partial \delta u_j}{\partial x_j}+ G \frac{\partial u_k}{\partial x_j}\frac{\partial \delta u_j}{\partial x_k}\right) \td V \\
&\qquad +\int_\Omega\sum_{\substack{k=1 \\ k \neq j}}^3  G \left(1+C_j^{\prime}\right)^2   \frac{\partial u_j}{\partial x_k}  \frac{\partial \delta u_j}{\partial x_k} \td V=0\,.
    \end{aligned}
\end{eqnarray}
In the current work, Eq.~\eqref{eq:weak form sec} for direction $j=\{1,2,3\}$ is to be solved at the PML domain along with the regular momentum balance in the non-PML domain given as:
\begin{eqnarray}
R(\tb u, \delta \tb u) = 
\int_{\Omega}\rho\ddot{\tb u} \cdot \delta \tb u\,\mathrm{d}{V}
+\int_{\Omega} \tb \sigma \colon \nabla \delta \tb u\,\mathrm{d}V =0\,.
\label{eq:weak_form_momentum}
\end{eqnarray}
Here, $\tb \sigma$ is the symmetric Cauchy's stress computed from the considered constitutive model.

\subsection{Spatial discretization}
In MPM, the volume integral of arbitrary physical quantities $\Phi(x)$ in the object domain can be approximated using the position vector $\tb x_p$ of particle and the particle's volume ${V}_p$ as follows\cite{sulsky,sulsky1994}:
\begin{eqnarray}
    \int_\Omega \Phi(\tb x^n) \td V \approx \sum_{p=1}^{n_p} \Phi(\tb x^n_p) V_p\,.
    \label{eq:mpm_discretization}
\end{eqnarray}
Meanwhile, the standard FE discretization can be employed to compute arbitrary physical quantities $\Phi(\tb x)$, along with their test functions, in each material point $p$. One can write:
\begin{eqnarray}
    \Phi(\tb x^n_p)= \sum_{I=1}^{n_n} \Phi(\tb x^n_I)N_I(\tb x^n_p)\,,
    \label{eq:fe_discretization}
\end{eqnarray}
where $N_I(\tb x^n_p)$ denotes the nodal basis function which is evaluated at the position of material point $x_p$. Here, we simplify the notation of the basis function $N_I(\tb x^n_p)\equiv N_{Ip}$ for brevity of the formulation. In addition, we defined $n_p$ and $n_n$ as the number of particles and nodes considered in spatial integration and interpolation.

\subsection{Mapping information from particle to grid and vice-versa}
\label{sec:mapping}

In the standard MPM methodology, the background nodes do not store any information as time steps advance. History-dependent information including kinematic variables, stresses, and material parameters, is stored in the material point. Therefore, it is necessary to transfer information from particles to the background nodes at the start of each time step. In the implicit MPM formulation, the nodal mass, velocity, and acceleration are initially computed as follows:
\begin{eqnarray}\label{eq:mapping}
m_I^n =\sum_{p=1}^{n_p} m_p N_{Ip}\,, \qquad 
\tb{v}_I^n =\frac{1}{m^n_I} \sum_{p=1}^{n_p} m_p \tb{v}_p^n N_{Ip}\,, \qquad
\tb{a}_I^n =\frac{1}{m^n_I}\sum_{p=1}^{n_p} m_p \tb{a}_p^n N_{Ip}\,.
\end{eqnarray}

In the current study, few modifications and additions are needed to map necessary information, particularly at the absorbing particle. First, the nodal mass needs to include the information of the damping function:\begin{eqnarray}\label{eq:mapping pml mass}
    \begin{aligned}
    m^n_{Ij} & =\sum_{p=1}^{n_p}\left(1+C^{\prime}_{pj}\right)^2 m_p N_{Ip}\,,
\end{aligned}
\end{eqnarray}
where the subscript $p$ in $C^{\prime}_{pj}$  represents the damping function defined for each absorbing particle. Notice that, the nodal mass expressed above becomes a direction-dependent nodal vector $\tb m_I^n$ instead of a scalar; its directional component is denoted by the subscript $j$. Using this nodal mass, the nodal displacement, velocity, and acceleration can be initially determined as follows for each direction $j$:
\begin{subequations}\label{eq:mapping pml prop}
    \begin{eqnarray}
u^n_{Ij}&=\ddfrac{1}{m^n_{Ij}}\displaystyle\sum_{p=1}^{n_p} \left(1+C_{pj}^{\prime}\right)^2m_p u^n_{pj} N_{Ip}\,, 
\\
v^n_{Ij} & =\ddfrac{1}{m^n_{Ij}}\displaystyle\sum_{p=1}^{n_p} \left(1+C_{pj}^{\prime}\right)^2 m_p v^n_{pj} N_{Ip}\,, \\
a^n_{Ij} & =\ddfrac{1}{m^n_{Ij}}\displaystyle\sum_{p=1}^{n_p} \left(1+C_{pj}^{\prime}\right)^2 m_p a^n_{pj} N_{Ip}\,.
\end{eqnarray}
\end{subequations}
Notice that, in the original formulation of implicit MPM, the nodal displacements do not need to be initialized. However, for some reasons discussed below, the nodal displacement field must also be mapped from particles to nodes in the proposed formulation. Moreover, it is worth emphasizing that the particle summation performed in Eqs.~\eqref{eq:mapping pml mass}-\eqref{eq:mapping pml prop} needs to generate continuous nodal field variables, which can be done by setting $C_{pj}=0$ for material points outside the PML domain.

At the end of each time step, when the solution to the discrete momentum balance is obtained, the interpolation of kinematic information from the background nodes to the particles is necessary. Here, we follow the modified FLIP update scheme \cite{brackbill1986flip} for implicit MPM as \cite{guilkey2003implicit, chandra2024stabilized}:
\begin{eqnarray}
\tb{u}_p^{n+1}&=&\sum_{I=1}^{n} N_{Ip}\tb{u}_I^{n+1}\,, \label{eq:disp_mapping} 
\\
\tb{a}_p^{n+1}&=&\sum_{I=1}^{n} N_{Ip}\tb{a}_I^{n+1}\,, \label{eq:acc_mapping}
\\
\tb{v}_p^{n+1}&=&\tb{v}_p^n+\frac{1}{2}\Delta t\left(\tb{a}_p^n+\tb{a}_p^{n+1}\right)\,. \label{eq:vel_mapping}
\end{eqnarray}

\subsection{Implicit displacement-based MPM implementation with PML}
In the current work, displacement-based MPM with an implicit time integration scheme is considered, where the previously discussed PML theory is implemented and discretized. Employing the Newmark-beta method, the discrete form of Eq.~\eqref{eq:weak form sec} can be constructed, yielding linear systems of equations in the following form, which is to be solved iteratively following the Newton-Raphson's method:
\begin{eqnarray}\label{eq:weal form for K and R}
    (\tb{K}_{IJ}^{\mathrm{tan}})^k \Delta \tb u_J^{k+1} =-\tb{R}_I^k\,.
\end{eqnarray}
Here, $\tb{R}$ represents the residual vector obtained from Eq.~\eqref{eq:weak form sec} after performing an MPM discretization (i.e.~Eq.~\eqref{eq:mpm_discretization}). Meanwhile, the superscript $k$ denotes Newton's iteration index. By invoking the arbitrariness of the nodal test functions, the residual vector can be expressed as follows:
\begin{eqnarray}
    \begin{aligned}
        {R}^k_{Ij} =& m^n_{Ij} \ddot{u}^k_{Ij}+
        \sum^{n_p}_{p=1}\Biggl( \sum_{J=1}^{n_n} (\lambda+2G)\frac{\partial N_{Ip}}{\partial x_j}\frac{\partial N_{Jp}}{\partial x_j} U^k_{Jj} \\
        & +
        \sum_{J=1}^{n_n}\sum_{\substack{m=1 \\ m \neq j}}^3\left\{\left(1+C_{pj}^{\prime}\right) \left(\lambda \frac{\partial N_{Ip}}{\partial x_m}\frac{\partial N_{Jp}}{\partial x_j}+
        G \frac{\partial N_{Ip}}{\partial x_j}\frac{\partial N_{Jp}}{\partial x_m}\right) U^k_{Jm} + \left. G\left(1+C_{pj}^{\prime}\right)^2 \frac{\partial N_{Ip}}{\partial x_m}  \frac{\partial N_{Jp}}{\partial x_m} U^k_{Jj} \right\}
       \right.\Biggl)V^k_p\,.
    \end{aligned}
\end{eqnarray}
In the expressions above and hereon, we differentiate three different discretized nodal displacement fields: the previous-step cumulative displacement $\tb u^n_I$, the incremental displacement $\tb u^{k}_I$, and the updated cumulative displacement $\tb U^k_I=\tb u^n_I+\tb u^k_I$. At the beginning of Newton's iteration, $k=0$, the incremental displacement is initiated as $\tb u^{(k=0)}_I=\tb 0$, which can be updated iteratively as $\tb u^{k+1}_I=\tb u^{k}_I+\Delta \tb u^{k+1}_I$. Once the convergence of the Newton-Raphson's algorithm is satisfied, the new displacement field can be assigned from the latest cumulative displacement as $\tb u^{n+1}_I = \tb U^{*}_I$, where the superscript $*$ indicates the last iteration at convergence.

In Eq.~\eqref{eq:weal form for K and R}, $\tb{K}^{\mathrm{tan}}$ denotes the tangent stiffness matrix. As the deformation of the PML element is significantly small, it is convenient to stick within the infinitesimal form of the tangent matrix, i.e.~the nonlinear geometric term is neglected for simplicity. The matrix can be divided into two sub-matrices:
\begin{eqnarray}
    \mathrm{\tb{K}^{tan}=\tb{K}^{mat}+\tb{K}^{dyn}}\,.
    \label{eq:tangent_stiffness}
\end{eqnarray}
The first sub-matrix is known as the material stiffness matrix and is expressed as follows:
\begin{eqnarray}
    \tb{K}^{\text{mat}}= \left[\begin{array}{ccc}
\tb{K}_{11}  & \cdots & \tb{K}_{1 N_n} \\
\vdots & \ddots & \vdots \\
\tb{K}_{N_n 1} & \cdots & \tb{K}_{N_n N_n} 
\end{array}\right]\,,
\end{eqnarray}
where $N_n$ indicates the total number of active nodes. For each block $\tb K_{IJ}$ corresponding to nodes $I$ and $J$ pair, we perform a direction-wise computation for $(m,n)=\{1,\,2,\,3\}$ as:
\begin{eqnarray}
   ({K}_{IJ})_{mn}=\begin{cases}
\displaystyle\sum^{n_p}_{p=1}\left(\frac{\partial N_{Ip}}{\partial x_m}\frac{\partial N_{Jp}}{\partial x_m} (2 G+\lambda)+\sum_{\substack{l=1 \\ l \neq m}}^3G \frac{\partial N_{Ip}}{\partial x_l} \frac{\partial N_{Jp}}{\partial x_l}(1+C_{pm}^\prime)^2\right)V_p\,, &(m=n)\,,\\
\displaystyle\sum^{n_p}_{p=1}\left((1+C_{pm}^\prime)\left(\frac{\partial N_{Ip}}{\partial x_m}\frac{\partial N_{Jp}}{\partial x_n}G+\frac{\partial N_{Ip}}{\partial x_n}\frac{\partial N_{Jp}}{\partial x_m} \lambda\right)\right)V_p\,,&(m \not= n)\,.
\end{cases}
\end{eqnarray}
The form above is a 3D generalization of the form proposed by \citet{chen2022} previously for 2D triangular elements. Moreover, the second sub-matrix in the right-hand side of Eq.~\eqref{eq:tangent_stiffness} is the dynamic component and can be computed as:
\begin{eqnarray}
     \tb{K}_{IJ}^{\mathrm{dyn}}=
\sum^{n_p}_{p=1} \frac{1}{\beta_N \Delta t^2}\mathbb{C}_p N_{Ip} m_p \tb I_{IJ}\,,
\end{eqnarray}
where matrix $\tb I$ is the identity matrix and $\mathbb{C}_p$ is a diagonal matrix containing the direction-dependent damping function, i.e.:
\begin{eqnarray}
\mathbb{C}_p=\begin{bmatrix}&
\left(1+C_{p1}^\prime\right)^2& 0& 0 \\& 0&\left(1+C_{p2}^\prime\right)^2&0\\
&0&0&(1+C_{p3}^\prime)^2
\end{bmatrix}\,.
\end{eqnarray}
In addition to that, the term $\beta_N$ denotes the variable of the Newmark-beta method. In this study, the non-dissipative choice of Newmark's parameters ($\gamma_N=0.5$ and $\beta_N=0.25$) is considered to minimize numerical damping from the time integration scheme.


\subsection{Further damping strategies}
\citet{chen2022} introduced the use of visco-elasticity and Rayleigh damping to the PML formulation to enhance the damping performance within the PML element. These two methods will also be added to the proposed MPM-PML formulation in the current work.

\subsubsection{Viscoelastic model with fractional derivative operators}
The viscoelastic behavior of materials can be introduced to the absorbing particles in the PML domain through a constitutive model that captures time-dependent deformation. In the present work, we utilized a model proposed by \citet{galucio} that employs the Gr\"unwald definition of fractional derivative operators. In one-dimensional settings, the time-dependent stress-strain relationship can be expressed as:
\begin{eqnarray}
\sigma(t)=E_0\left(\left(1+c \frac{\left(E_{\infty}-E_0\right)}{E_0}\right) \varepsilon(t)+c \frac{E_{\infty}}{E_0} \sum_{q=1}^{N_t}\left(A_{q+1}\, \bar{\varepsilon}(t-q \Delta t)\right)\right)\,,
\label{eq:1d_stress_visco}
\end{eqnarray}
where $\sigma(t)$ denotes the 1D stress, defined based on relaxed elastic modulus $E_0$, non-relaxed elastic modulus $E_\infty$, strain \(\varepsilon(t)\), and internal strain $\bar{\varepsilon}(t)$. In the above expressions, $N_t$ dictates how many time steps of strain history are to be considered. Furthermore, $A_q$ is the term involving the Grünwald coefficient, which incorporates the memory effect inherent in viscoelastic materials \cite{galucio}, whereas $c$ is a coefficient that balances the fractional derivative's impact ($0 \leq c \leq 1$). The Grünwald coefficient and the coefficient $c$ are defined as:
\begin{eqnarray}
    A_{q+1} &=& \frac{q-\alpha-1}{q} A_q\,, \qquad \mathrm{where}\quad A_1 = 1\,, \\
c&=&\frac{(\tau)^\alpha}{(\tau)^\alpha+(\Delta t)^\alpha}\,,
\end{eqnarray}
where $\alpha$ is the fractional order ($0 \leq \alpha \leq1$). Here, $\tau$ and $\Delta t$ are the relaxation time and the time step, respectively.

In Eq.~\eqref{eq:1d_stress_visco}, the internal strain $\bar{\varepsilon}(t)$ is defined as:
\begin{eqnarray}\label{eq:internal strain}
    \bar{\varepsilon}(t) = \left((1-c)\frac{E_\infty - E_0}{E_\infty} \right) \varepsilon(t) - c \sum_{q=1}^{N_t}\Bigl( {A}_{q+1}\overline{\varepsilon}(t-q \Delta t)\Bigr)\,.
\end{eqnarray}
Here, the following relationship between strain and displacement is valid:
\begin{eqnarray}
    \bar{\varepsilon}(t)=B \bar{u}_v(t)\,,
\end{eqnarray}
where $B$ represents the spatial differentiation operator and $\bar{u}_v(t)$ denotes the displacement associated with the defined internal strain function. From Eq.~\eqref{eq:internal strain}, $\bar{u}_v(t)$ is defined as follows:
\begin{eqnarray}\label{eq:displacement dif}
\bar{u}_v(t)&=&(1-c) \frac{\left(E_{\infty}-E_0\right)}{E_{\infty}} u(t)-c \sum_{q=1}^{N_t}\left(A_{q+1} \bar{u}_v(t-q \Delta t)\right)\,,\\
\bar{u}_v(0)&=&(1-c) \frac{\left(E_{\infty}-E_0\right)}{E_{\infty}} u(0)\,.
\end{eqnarray}
Notice that at a given time $t$, the internal displacement $\bar{u}_v(t)$ can be computed from the regular displacement field $u(t)$ and some components of the internal displacements obtained in the previous time steps, $t-q\Delta t$. The extension of this viscoelastic model to 3D is straightforward. The detailed implementation of this model to MPM will be presented later in \Cref{subsubsec:implementation_damping}.


\subsubsection{Rayleigh damping within PML domain}
In the current work, we consider an additional Rayleigh damping to be applied within the PML domain. Here, the damping matrix can be written as a linear function of the modified lumped mass matrix, $\tb M$, where its components can be evaluated following Eq.~\eqref{eq:mapping pml mass}. One can write:
\begin{eqnarray}\label{eq:rayleigh}
    \tb{C}^M=\alpha_M \tb{M}\,,
\end{eqnarray}
where $\alpha_M$ indicates the mass-proportional damping coefficient. In MPM, the local Rayleigh damping force in each node can be computed considering the nodal lumping of the nodal mass as follows:
\begin{eqnarray}
    \tb{f}^c_I = \tb C^M_I \odot \dot{\tb u}_{I} = \alpha_M \tb m^n_I \odot \dot{\tb u}_{I}\,,
\end{eqnarray}
where $\odot$ denotes the component-wise multiplication of the nodal mass and the nodal velocity vectors.

\subsubsection{Implementation of damping terms within PML domain}
\label{subsubsec:implementation_damping}

In the following section, we rewrite the new residual vector and tangential stiffness formulations specific to the PML elements. Here, we denote the modified residual vector as $\widetilde{\tb{R}}$, where it can be computed at each Newton's iteration $k$ as follows:
\begin{eqnarray}\label{eq:final residual}
    \begin{aligned}
        \widetilde{R}^k_{Ij} =&m^n_{Ij} \ddot{u}^k_{Ij}+C_{Ij}^M \dot{u}^k_{Ij}+ c \frac{E_{\infty}}{E_0}\left(\sum_{J=1}^{n_n} \tb K^{\mathrm{mat}}_{IJ} \sum_{q=1}^{N_t}\left(A_{q+1}\right) {(\tb{\bar{u}}^{n-q}_{vJ})}\right)_j\\
       &+(1+c) \frac{\left(E_{\infty}-E_0\right)}{E_{\infty}} \sum^{n_p}_{p=1}\Biggl( \sum_{J=1}^{n_n} (\lambda+2G)\frac{\partial N_{Ip}}{\partial x_j}\frac{\partial N_{Jp}}{\partial x_j} U^k_{Jj}\\
       & \quad+
        \sum_{J=1}^{n_n} \sum_{\substack{m=1 \\ m \neq j}}^3\left\{\left(1+C_{pj}^{\prime}\right) \left(\lambda \frac{\partial N_{Ip}}{\partial x_m}\frac{\partial N_{Jp}}{\partial x_j}+
        G \frac{\partial N_{Ip}}{\partial x_j}\frac{\partial N_{Jp}}{\partial x_m}\right)U^k_{Jm}
       \right. \\
        & \qquad\left. + G\left(1+C_{pj}^{\prime}\right)^2 \frac{\partial N_{Ip}}{\partial x_m}  \frac{\partial N_{Jp}}{\partial x_m} U^k_{Jj} \right\} \Biggl)V^k_p\,.
    \end{aligned}
\end{eqnarray}
In the above expression, $\tb{\bar{u}}^{n}_{vI}$ can be computed as follows:
\begin{eqnarray}\label{visco-particle-disp}
    \tb{\bar{u}}^{n}_{vI} &=& \ddfrac{1}{m^n_{Ij}}\displaystyle\sum_{p=1}^{n_p} \left(1+C_{pj}^{\prime}\right)^2m_p \tb{\bar{u}}^n_{vp} N_{Ip}\,,\\
    \tb{\bar{u}}^n_{vp}&=&(1-c) \frac{\left(E_{\infty}-E_0\right)}{E_{\infty}} \tb{u}^n_p-c \sum_{q=1}^{N_t}\left(A_{q+1} \tb{\bar{u}}_{vp}^{n-q}\right)\,.
\end{eqnarray}
In the current work, the parameter $N_t$ is fixed to be 4 to balance both computational costs to store the internal displacement variables and performance of the visco-elastic model. With $N_t=4$, the value of Gr\"unwald coefficient $A_5$ is determined to be $-4.2\times 10^{-3}$, which is considered to be sufficiently small and does not anymore compromise the accuracy and reliability of the model.

For the left-hand side terms, we denote the modified tangential stiffness matrix component as $\widetilde{\tb{K}}^{\mathrm{mat}}$ and $\widetilde{\tb{K}}^{\mathrm{dyn}}$, where they can be computed respectively as follows:
\begin{eqnarray}\label{eq:final Kmat}
    \widetilde{\tb K}^{\text {mat}}&=& \left(1+c \frac{E_{\infty}-E_0}{E_0}\right) \tb K^{\text {mat}}\,,\\
\label{eq:final dyn}
     \widetilde{\tb K}_{IJ}^{\mathrm{dyn}}&=& \sum^{n_p}_{p=1}
\left(\frac{1}{\beta_N \Delta t^2}+\frac{\alpha_M \gamma_N}{\beta_N \Delta t } \right)\mathbb{C}_p N_{Ip} m_p \tb I_{IJ}\,.
\end{eqnarray}
Finally, by substituting Eqs.~\eqref{eq:final residual}-\eqref{eq:final dyn} into Eq.~\eqref{eq:weal form for K and R}, the linear systems of equations can be solved using classical iterative algorithms, e.g.~\cite{implicit1,implicit2}. 

\subsection{Summary of computational procedure}
For simulations involving elasto-plastic confinement-dependent material properties, the initial stress field is often required before performing any subsequent analysis. To obtain the in-situ stress fields, a geo-static step is initially performed. In this step, the absorbing particles are treated as general material points and their attenuation with respect to gravity is not considered. We denote the initial static stress as $\tb{\sigma}_{\text{init}}$, which can be connected to gravitational load particle-wise as follows during the static condition:
\begin{eqnarray}\label{eq:stress initial}
    \tb{B}_{Ip} \tb{\sigma}_{p,\text{init}} = -\rho_p \tb b_p N_{Ip}\,,
\end{eqnarray}
where $\tb{B}_I$ is the deformation matrix related to each node $I$ in the element and $\tb b$ indicates the gravitational acceleration. The geo-static step is also performed to achieve the initial $K_0$ condition commonly considered in geotechnical analysis. 

The initial stress field obtained in the geo-static step is then appended to the balance equation along with setting the initial displacement to zero. The residual vector component given in Eq.~\eqref{eq:final residual} can be rewritten as follows:
\begin{eqnarray}\label{eq:final residual gravity}
    \begin{aligned}
        \widetilde{R}^k_{Ij} =&m^n_{Ij} \ddot{u}^k_{Ij}+C_{Ij}^M \dot{u}^k_{Ij}+ c \frac{E_{\infty}}{E_0}\left(\sum_{J=1}^{n_n} \tb K_{IJ}^{\mathrm{mat}} \sum_{q=1}^{N_t}\left(A_{q+1}\right) {(\tb{\bar{u}}^{n-q}_{vJ})}\right)_j\\
       &+(1+c) \frac{\left(E_{\infty}-E_0\right)}{E_{\infty}} \sum^{n_p}_{p=1}\Biggl( \sum_{J=1}^{n_n} (\lambda+2G)\frac{\partial N_{Ip}}{\partial x_j}\frac{\partial N_{Jp}}{\partial x_j} U^k_{Jj}\\
       &\quad+\sum_{J=1}^{n_n}\sum_{\substack{m=1 \\ m \neq j}}^3\left\{\left(1+C_{pj}^{\prime}\right) \left(\lambda \frac{\partial N_{Ip}}{\partial x_m}\frac{\partial N_{Jp}}{\partial x_j}+
        G \frac{\partial N_{Ip}}{\partial x_j}\frac{\partial N_{Jp}}{\partial x_m}\right)U^k_{Jm}
       \right. \\
        & \qquad+\left. G\left(1+C_{pj}^{\prime}\right)^2 \frac{\partial N_{Ip}}{\partial x_m}  \frac{\partial N_{Jp}}{\partial x_m} U^k_{Jj} \right\}\Biggl)V^k_p- \sum^{n_p}_{p=1}\Biggl(\rho b^n_{pj} N_{Ip} + (\tb{B}_{Ip} \tb{\sigma}_{p,\text{init}})_j \Biggl)V^k_p\,.
    \end{aligned}
\end{eqnarray}
When the body force is only gravity, Eq.~\eqref{eq:stress initial} shows that the initial stress field is balanced by gravity, so the damping of the absorbing particles does not act on the gravity load and is only effective during the dynamic step. During material deformation, the change in the stress field due to a change in geometry results in a displacement field that satisfies the dynamic equilibrium equation. In other words, any difference in stress from the initial stress is represented by a displacement field. However, this displacement may be discontinuous between the PML domain and the elastic base due to differences in the stiffness matrix. It is worth emphasizing that the volumetric strain caused by the change in geometry is so small that this displacement discontinuity is also remarkably small.

\begin{figure}[h!]
    \centering
    \includegraphics[width=\linewidth]{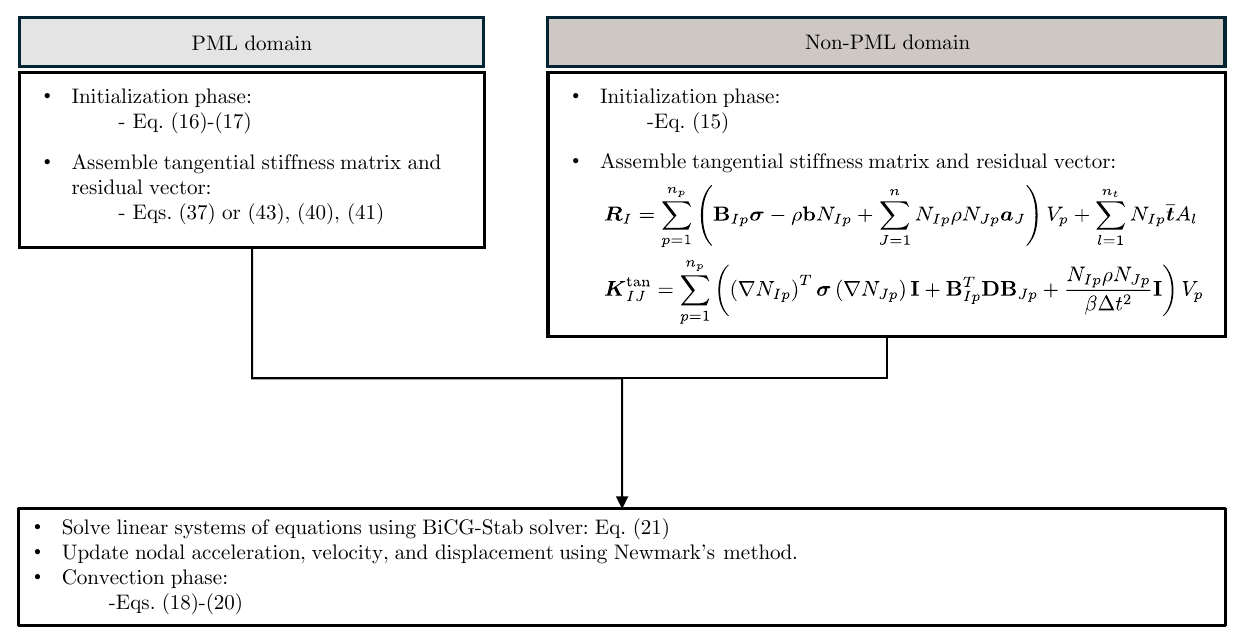}
    \caption{Summary of computational procedure of the dynamic step.}
    \label{fig:algo}
\end{figure}

\begin{figure}[h!]
    \centering
    \includegraphics[width=0.85\linewidth]{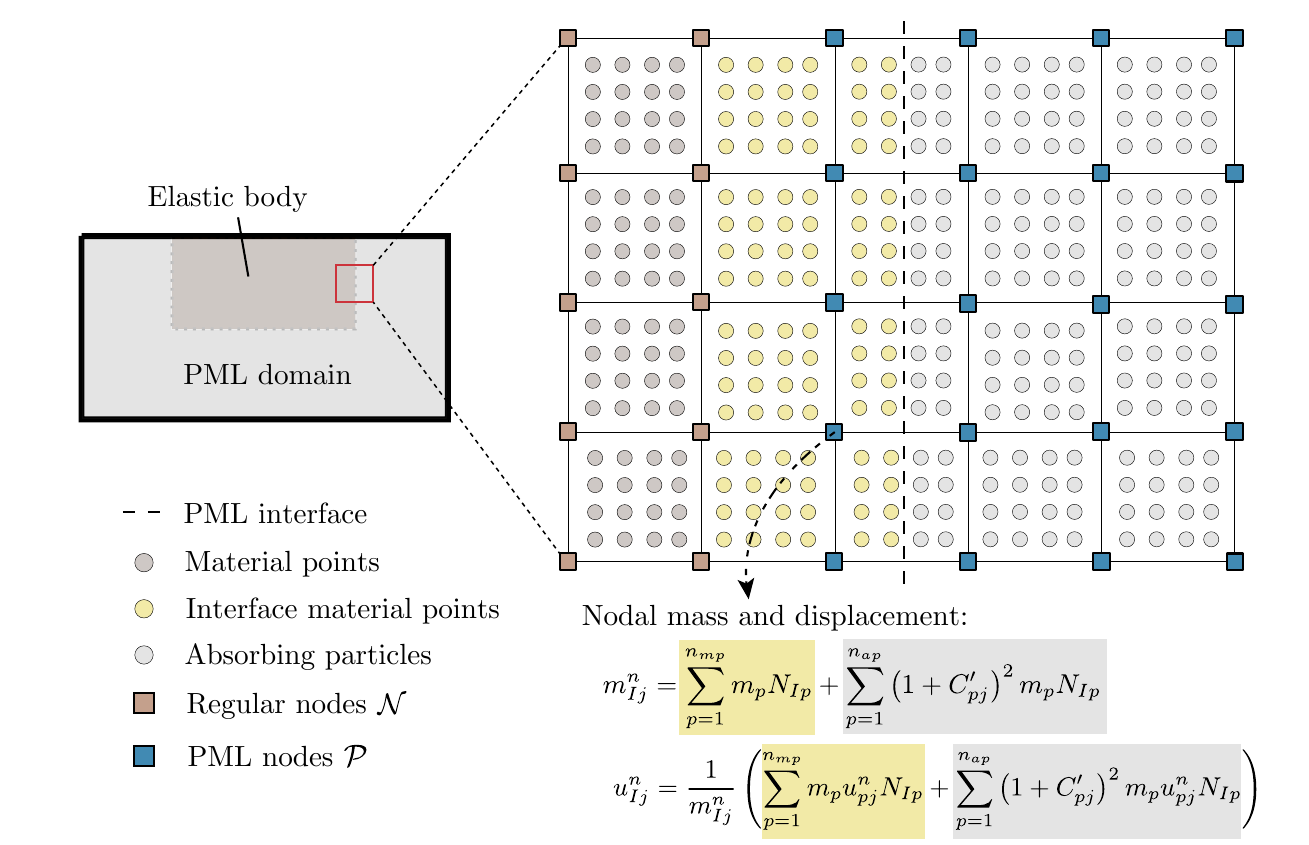}
    \caption{Mapping mass and displacement fields from both material points and absorbing particles near the PML interface. Here, $n_{mp}$ and $n_{ap}$ denote the number of material points and absorbing particles associated with a given PML node near the interface, respectively. The interface material points can be identified as material points with at least one connectivity marked as PML nodes, i.e.~$I \in \mathcal{P}$. Here, for simplicity of the illustration, a standard "local" quadrilateral element is considered. However, the extension to higher-order "non-local" basis functions should follow an analogous approach.}
    \label{fig:interface}
\end{figure}

Subsequently, the algorithm during the dynamic step is summarized in Fig.~\ref{fig:algo} together with the equations needed at the non-PML domain. The detailed formulation of the non-PML domain follows the standard displacement-based implicit MPM as suggested by \citet{guilkey2003implicit}. \revision{At the beginning of each time step, the nodes associated with the absorbing particles are first identified; we denote $\mathcal{P}$ as the PML node set and $\mathcal{N}$ as the regular nodes not associated with the PML domain. Once these nodes are identified, the particle-to-grid mapping should be performed independently, as described in \Cref{sec:mapping}. Care must be taken when mapping mass information at the boundary between PML and non-PML domains; specifically, the nodal mass is now a vector in each node in $\mathcal{P}$, rather than a scalar. This can be achieved by specializing the mapping function for material points located in cells adjacent to the PML domain, which we refer to as \textit{interface material points} (highlighted in yellow in Fig.~\ref{fig:interface}). Since nodal displacement information is required at PML nodes to compute internal forces (cf.~Eq.~\eqref{eq:final residual} and \eqref{eq:final residual gravity}), the displacements of interface material points should be mapped to PML nodes, along with other kinematic quantities such as velocity and acceleration. Given that the current PML framework assumes a displacement-based formulation, consistent with the implicit MPM scheme, the external and internal forces, as well as the assembly of the tangent stiffness matrix, can be handled straightforwardly using the standard FE procedures, incorporating contributions from both the interface material points and the absorbing boundaries at the interface PML nodes.} 

In the current work, dynamic loadings are applied by imposing Neumann traction at the particles near the Neumann boundary at the non-PML domain. Once the global tangent matrix and residual vectors are assembled, the linear systems of equations, Eq.~\eqref{eq:weal form for K and R}, are solved using a biconjugate gradient stabilized (BiCGStab) solver; the current work utilized the solver available within the Eigen solver library \cite{eigenweb}. Once the Newton-Raphson's iterations converged, the updated kinematic variables are advected back to the particles following Eqs.~\eqref{eq:disp_mapping}-\eqref{eq:vel_mapping}.

\section{Numerical examples}
\label{sec:num_examples}

\subsection{Two‑dimensional elastic soil under point-source impact loading}\label{elastic_2d}
\begin{figure}[h!]
  \begin{minipage}[b]{0.45\textwidth}
    \centering
    \includegraphics[width=\textwidth]{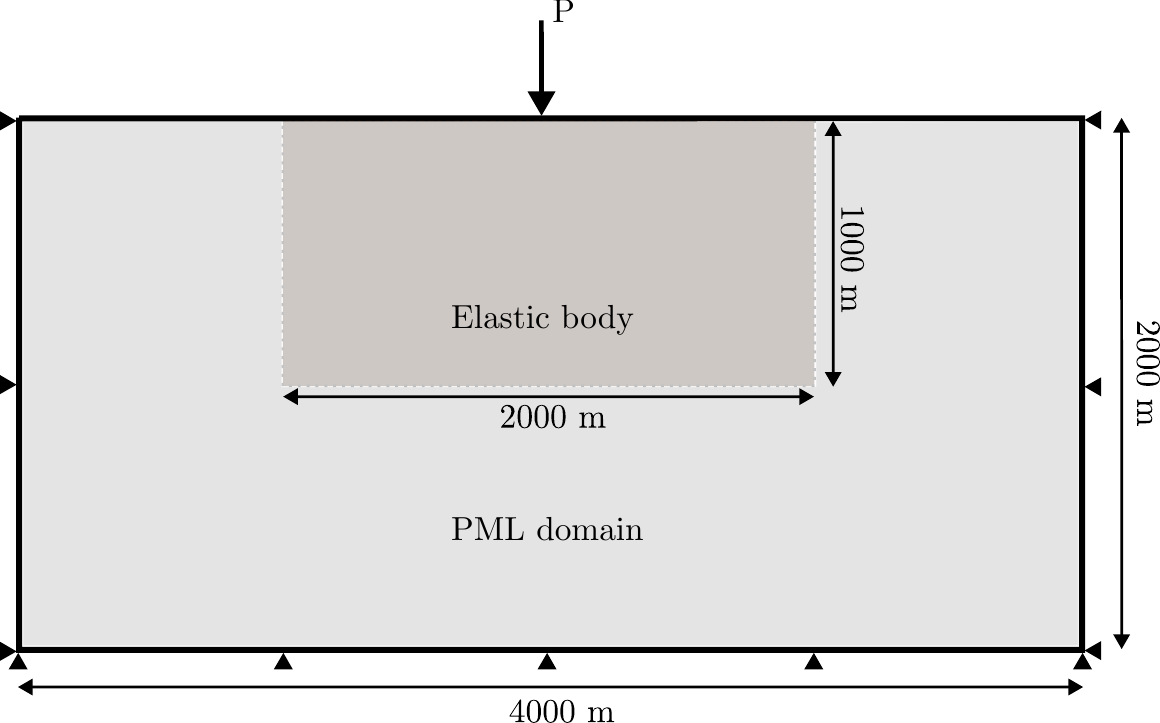}
    \subcaption{Computational domain}
  \end{minipage}
  \hspace{0.04\columnwidth}
  \begin{minipage}[b]{0.41\textwidth}
    \centering
    \includegraphics[width=\textwidth]{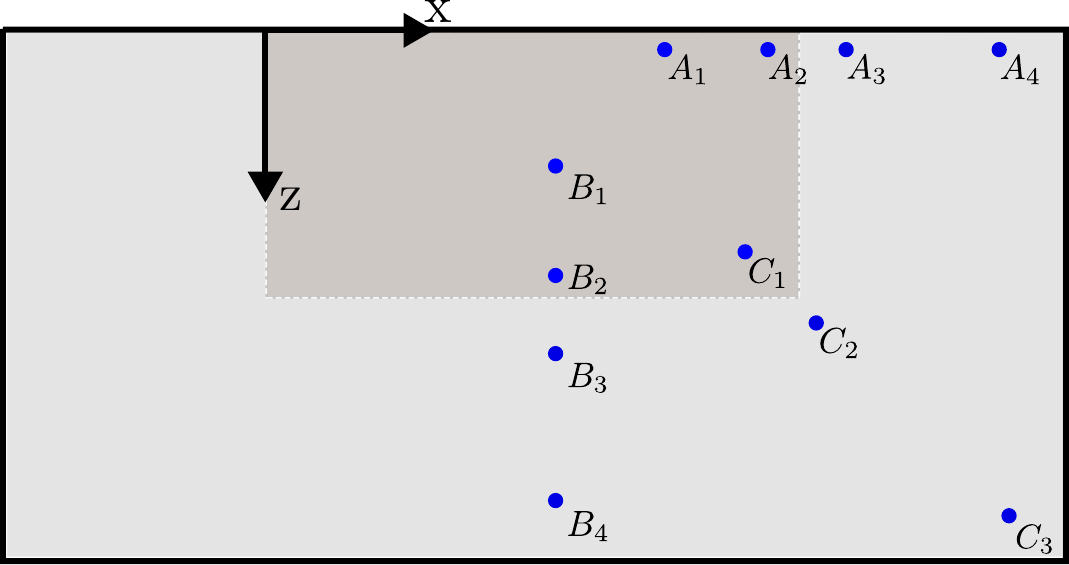}
    \vspace{0.05cm}
    \subcaption{Receiver points}
  \end{minipage}
  \caption{2D elastic soil under point loading: model geometry and boundary conditions.}
  \label{fig:Lame problem}
\end{figure}

In this section, we perform a numerical analysis of a 2D semi-infinite plane problem assuming an elastic body. In this problem, we simulate the propagation of elastic waves in an isotropic, homogeneous half-space. This model employs a directional force point source applied to the upper free surface of the domain. To realize the semi-infinite plane, absorbing particles are placed surrounding the target domain to absorb outgoing waves and suppress reflected waves. Fig.~\ref{fig:Lame problem} depicts the simulation domain, boundary conditions, and the location of receiver points. The coordinates for the receiver points are given as: \revision{A1(1525,65), A2(1805,65), A3(2205,65), A4(2795,65), B1(1125,575), B2(1125,805), B3(1125,1205), B4(1125,1795), C1(1725,725), C2(1105,1105), C3(1865,1865)}, all measured from the coordinate origin located at the \revision{top left corner of the elastic body}. 

The background cell is set as 20 m $\times$ 20 m, where initially each cell contains four particles, yielding a total of 80,000 particles: 20,000 material points and 60,000 absorbing particles. Young’s modulus $E$, Poisson’s ratio $\nu$, and mass density $\rho$ are set as 2 GPa, 0.25, and 2000 $\mathrm{kg/m}^3$, respectively. The time increment is set as \revision{$\Delta t=0.01$ s}. For the PML damping function's parameters, $\alpha_j$ (the maximum values of ${C}_j^{\prime}$) and $\beta$ (polynomial power of ${C}_j^{\prime}$), we adopted the values of 4 and 1 respectively, as they were suggested as the optimal values in a similar case study conducted by \citet{chen2022} \revision{(we further confirm these values through a sensitivity analysis in \ref{app:pml_param_sens_anal})}. The relaxed Young's modulus and the fractional order for the visco-elastic model are set to be $E_0 = 0.99 E = 1.98$ GPa and $\alpha = 0.95$, respectively. Meanwhile, the relaxation time is set to \revision{$\tau=2\Delta t=0.02$ s}. To achieve a greater damping effect, we have deliberately set the Rayleigh damping factor $\alpha_M$ to 1.0, which is higher than the typical values used for modeling numerical damping. Note that, in the current model and subsequent models, the visco-elastic damping and Rayleigh damping are only applied within the PML domain, not the main elastic (or elasto-plastic) domain.

\begin{figure}[h!]
    \centering
     \begin{subfigure}[b]{0.32\textwidth}
         \centering
         \includegraphics[width=\textwidth]{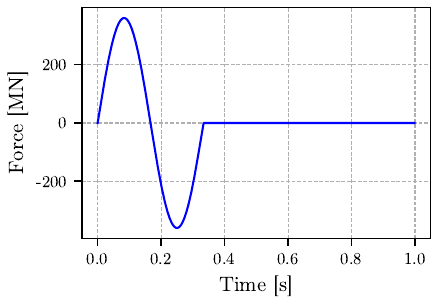}
         \caption{Test 1}
         \label{fig:Input wave1}
     \end{subfigure}
     \hspace{0.1cm}
     \begin{subfigure}[b]{0.32\textwidth}
         \centering
        \includegraphics[width=\linewidth]{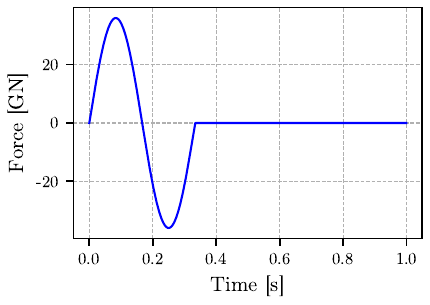}
        \caption{Test 2}
        \label{fig:Input wave2}
     \end{subfigure}
     \hspace{0.1cm}
      \begin{subfigure}[b]{0.32\textwidth}
         \centering
        \includegraphics[width=\linewidth]{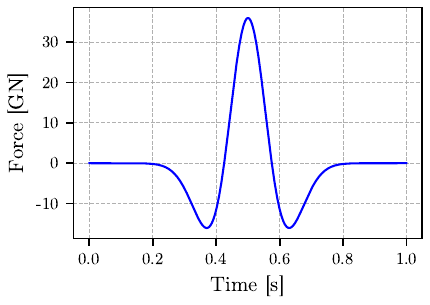}
    \caption{Test 3}
    \label{fig:Input wave3}
     \end{subfigure}
    \caption{2D elastic soil under point loading: input impulse loadings for three different tests.}
    \label{fig:3.1_input_waves}
\end{figure}

Using the same geometry and conditions, we conducted three validation tests, each with different loading conditions (cf.~Fig.~\ref{fig:3.1_input_waves}). Test 1 involves a \revision{small} amplitude \revision{sinusoidal pulse force}, as illustrated in Fig.~\ref{fig:Input wave1}, while Test 2 is considered to specifically analyze situations involving large deformation with a large amplitude of impulse force ($\times 100$ of the load of Test 1), as illustrated in Fig.~\ref{fig:Input wave2}. For Test 3, the input wave is \revision{a Ricker wavelet with a central frequency of 3 Hz} as illustrated in Fig.~\ref{fig:Input wave3}. Since MPM is often employed to model large-deformation problems, this analysis aims to assess the performance of the proposed implementation for such conditions. To illustrate the improved performance, we compared the obtained numerical results with the same simulations performed with the standard implicit MPM scheme without absorbing particles. \revision{At the same time, a reference analysis with an extended simulation domain was conducted. An elastic buffer zone of 7000 m was added around the target domain to ensure that waves refracted by the boundaries would not reach the receiver points within the selected time interval. It is important to note that different choices of PML parameters may lead to slightly different wave propagation results and absorption performance. Therefore, we conducted an additional sensitivity analysis of these parameters in \ref{app:pml_param_sens_anal}, using Test 2 as a case study.}

\begin{figure}[h!]
    \centering
    \begin{subfigure}{0.32\linewidth}
        \includegraphics[width=\linewidth]{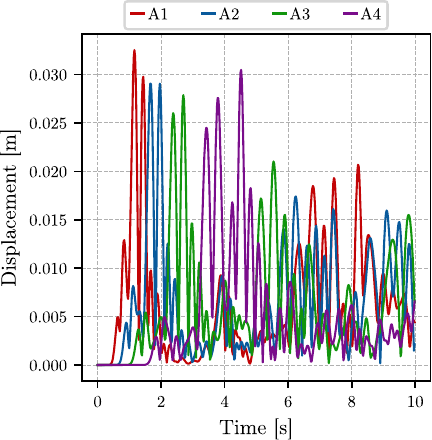}
    \end{subfigure}
    \begin{subfigure}{0.32\linewidth}
        \includegraphics[width=\linewidth]{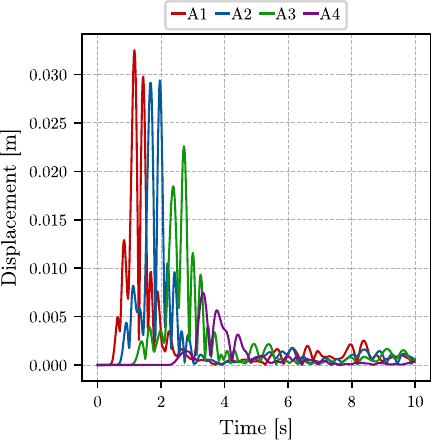}
    \end{subfigure}
        \begin{subfigure}{0.32\linewidth}
        \includegraphics[width=\linewidth]{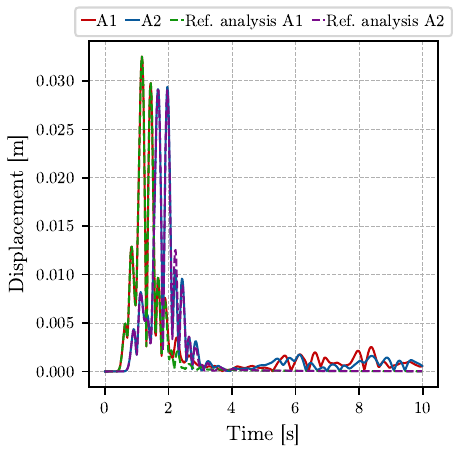}
    \end{subfigure}
    \\
    \begin{subfigure}{0.32\textwidth}
        \includegraphics[width=\linewidth]{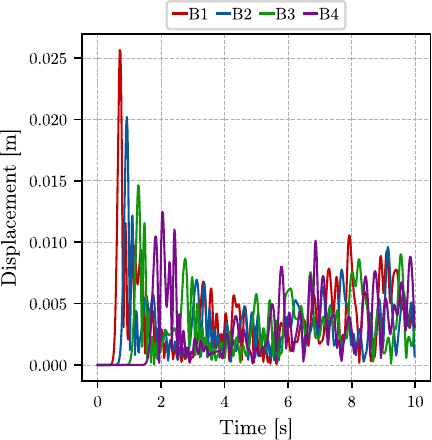}
    \end{subfigure}
    \begin{subfigure}{0.32\textwidth}
        \includegraphics[width=\linewidth]{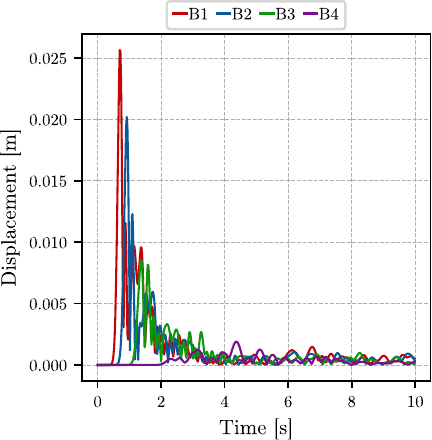}
    \end{subfigure}
        \begin{subfigure}{0.32\textwidth}
        \includegraphics[width=\linewidth]{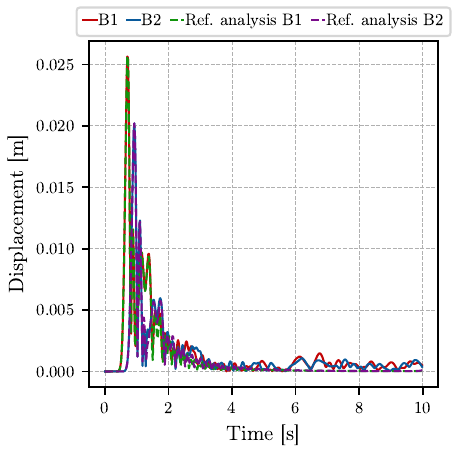}
    \end{subfigure}
    \\
    \begin{subfigure}{0.32\textwidth}
        \includegraphics[width=\linewidth]{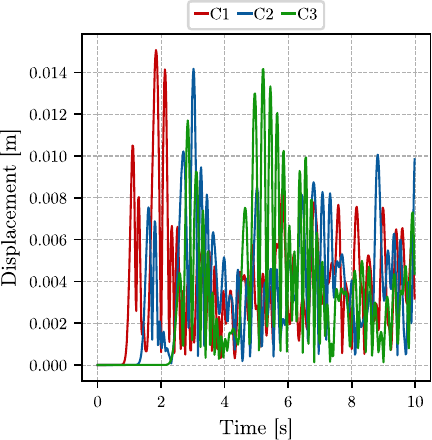}
        \caption{Standard implicit MPM}
    \end{subfigure}
    \begin{subfigure}{0.32\textwidth}
        \includegraphics[width=\linewidth]{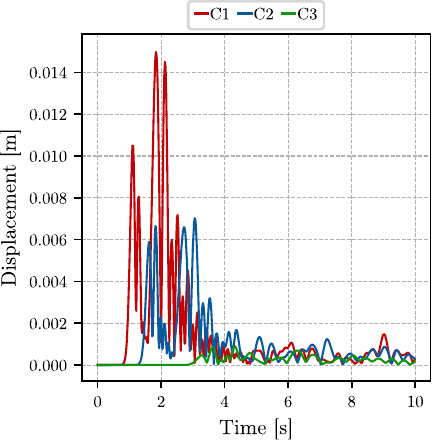}

        \caption{Proposed method}
    \end{subfigure}
    \begin{subfigure}{0.32\textwidth}
        \includegraphics[width=\linewidth]{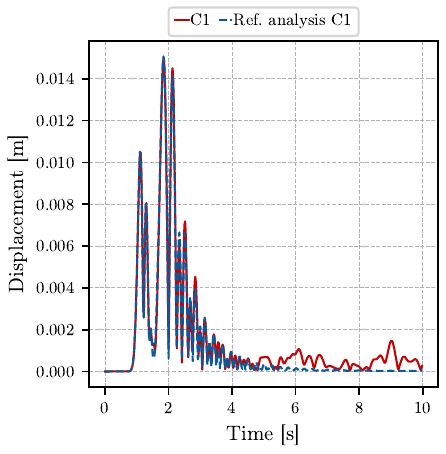}

        \caption{Proposed method with reference}
    \end{subfigure}
    \caption{2D elastic soil under point loading -- Test 1: displacement magnitudes comparisons between the standard implicit MPM scheme, the proposed method, and the reference simulation.}
    \label{fig:test1 graph}
\end{figure}

Fig.~\ref{fig:test1 graph} plots the displacement magnitudes for Test 1 measured at the aforementioned receiver points, with a comparison to \revision{the reference solution and} results without absorbing particles  The analyses of experiments A1-A4 demonstrate that our proposed methodology effectively attenuates horizontally propagating waves. Similarly, observations from B1-B4 reveal that vertically propagating waves are also mitigated by our approach. The data from C1-C3 further elucidate that our method attenuates waves traversing in both horizontal and vertical planes. Notably, findings from C3 illustrate that our methodology significantly diminishes the amplitude of diagonally propagating waves. The ratio of the maximum amplitude in comparison to the standard implicit MPM after 5 seconds averaged about \revision{10\%} for A1-A4, \revision{12.1\%} for B1-B4, and \revision{11.96\%} for C1-C3. From the given plots, we can observe that the proposed method can absorb outgoing waves quite well, limiting any significant reflection waves from returning to the target domain. \revision{In comparison to the reference solution, the PML method performs relatively well. However, minor fluctuation errors are still observed in the later stages of the recording, $t\geq3$ s. Several reasons that may cause this error and improvements to alleviate these errors will be discussed with the other two tests in later paragraphs.}

\begin{figure}[h!]
    \centering
    \begin{subfigure}{0.32\textwidth}
        \includegraphics[width=\linewidth]{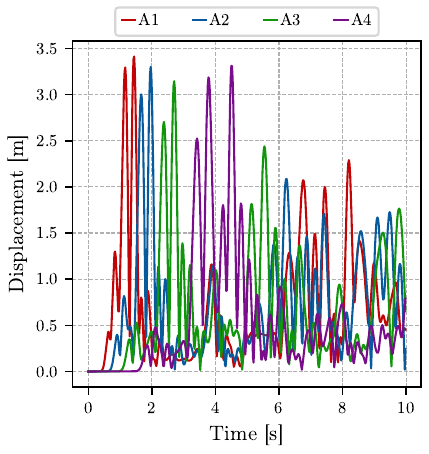}
    \end{subfigure}
    \begin{subfigure}{0.32\textwidth}
        \includegraphics[width=\linewidth]{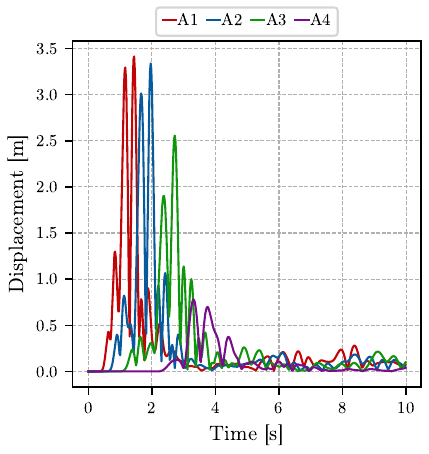}
    \end{subfigure}
    \begin{subfigure}{0.33\textwidth}
        \includegraphics[width=\linewidth]{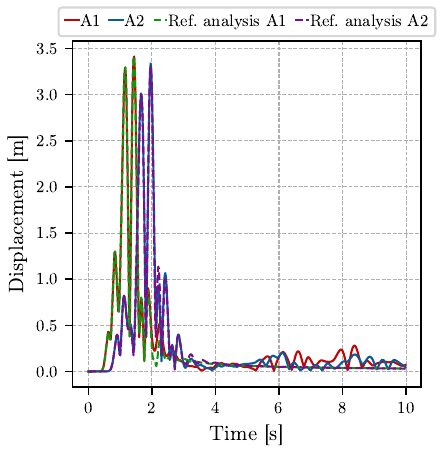}
    \end{subfigure}
    \\
    \begin{subfigure}{0.32\textwidth}
        \includegraphics[width=\linewidth]{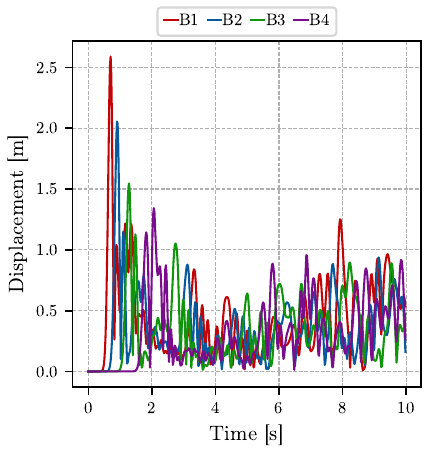}
    \end{subfigure}
    \begin{subfigure}{0.32\textwidth}
        \includegraphics[width=\linewidth]{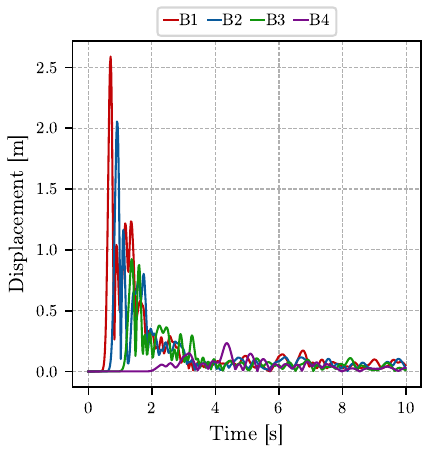}
    \end{subfigure}
    \begin{subfigure}{0.33\textwidth}
        \includegraphics[width=\linewidth]{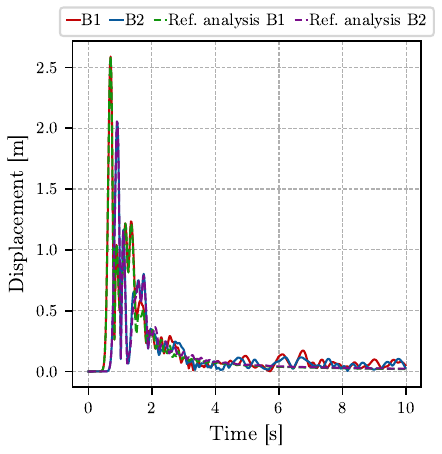}
    \end{subfigure}
    \\
    \begin{subfigure}{0.32\textwidth}
        \includegraphics[width=\linewidth]{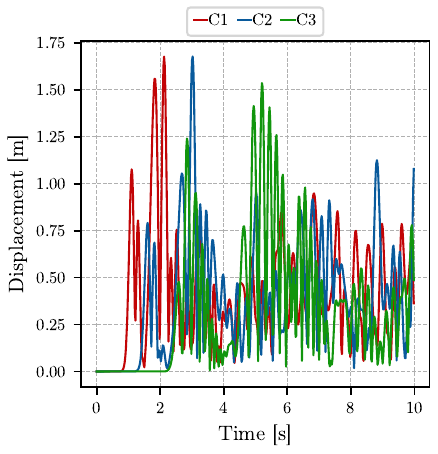}
        \caption{Standard implicit MPM}
    \end{subfigure}
    \begin{subfigure}{0.32\textwidth}
        \includegraphics[width=\linewidth]{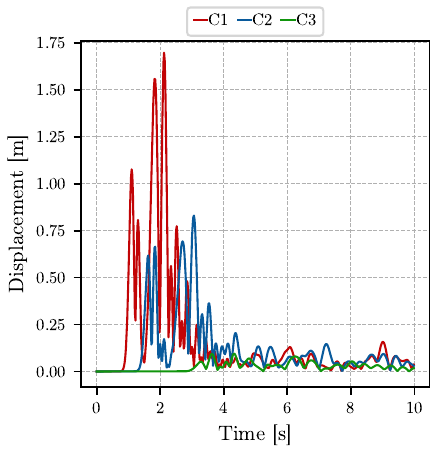}
        \caption{Proposed method}
    \end{subfigure}
    \begin{subfigure}{0.32\textwidth}
        \includegraphics[width=\linewidth]{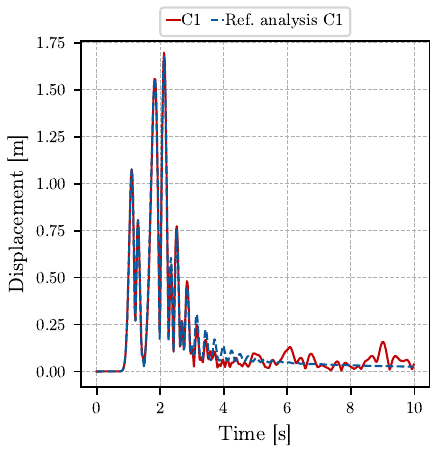}
        \caption{Proposed method with reference}
    \end{subfigure}
    \caption{2D elastic soil under point loading -- Test 2: displacement magnitudes comparisons between the standard implicit MPM scheme, the proposed method, and the reference simulation.}
    \label{fig:test2 graph}
\end{figure}

\begin{figure}[h!]
    \centering
  \includegraphics[width=\linewidth]{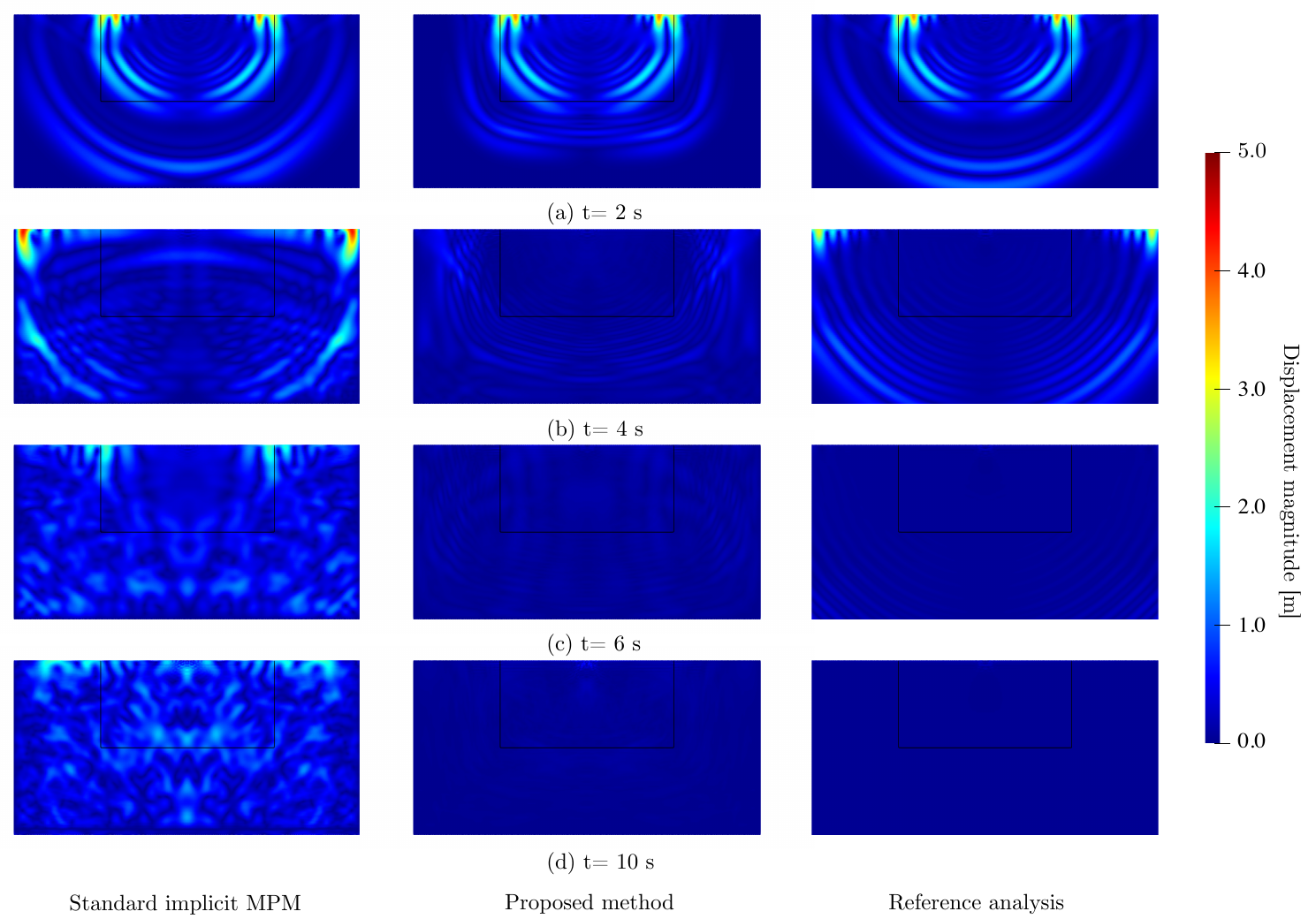}  
    \caption{2D elastic soil under point loading -- Test 2: displacement magnitude contours obtained by (left) standard implicit MPM, (center) the proposed method, and (right) the reference analysis with an enlarged domain at $t$ = 2.0, 4.0, 6.0, and 10.0 s.}
        \label{fig:test2 screenshot}
\end{figure}

Fig.~\ref{fig:test2 graph} illustrates the displacement magnitudes for Test 2. Similar to the results obtained in Test 1, all observation points A to C demonstrate that the proposed method attenuates reflected waves in vertical, horizontal, and diagonal directions. The ratio of the maximum amplitude \revision{compared to the standard implicit MPM results} after 5 seconds is about \revision{7.24\%} for A1-A4, \revision{9.81\%} for B1-B4, and \revision{7.69\%} for C1-C3. From this test, it is evident that even though the magnitude of the maximum displacement has increased by more than 100 times compared to those in Test 1, the proposed method remains effective with a minor reduction in damping capacity. In addition, Fig.~\ref{fig:test2 screenshot} presents the magnitude of displacement contours for Test 2 captured at a few different time snapshots. As highlighted in the figures in the \revision{center} column, despite the relatively coarse mesh, the proposed implementation is able to absorb most of the outgoing waves even with comparatively large displacement waves. 





\begin{figure}[h!]
    \centering
    \begin{subfigure}{0.32\textwidth}
        \includegraphics[width=\linewidth]{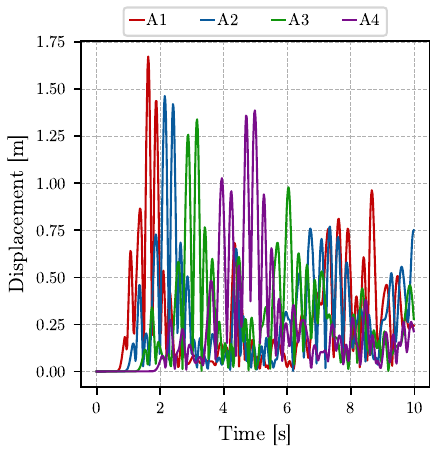}

    \end{subfigure}
    \begin{subfigure}{0.32\textwidth}
        \includegraphics[width=\linewidth]{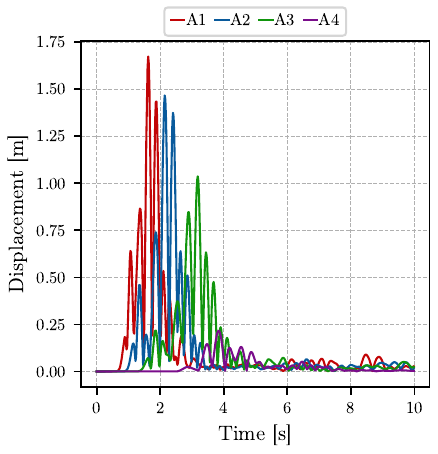}

    \end{subfigure}
        \begin{subfigure}{0.33\textwidth}
        \includegraphics[width=\linewidth]{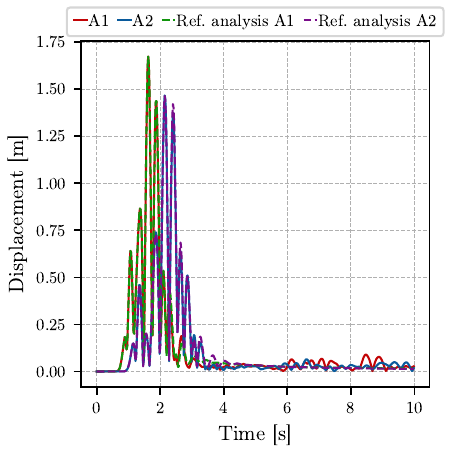}

    \end{subfigure}
    \\
    \begin{subfigure}{0.32\textwidth}
        \includegraphics[width=\linewidth]{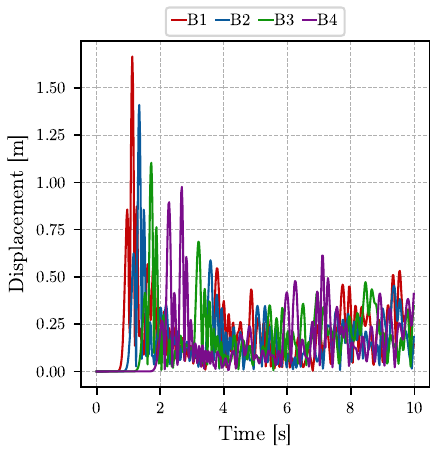}

    \end{subfigure}
    \begin{subfigure}{0.32\textwidth}
        \includegraphics[width=\linewidth]{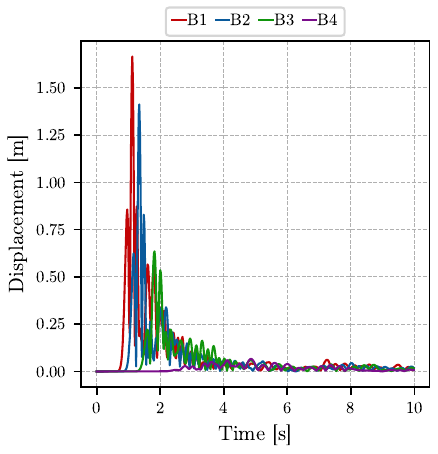}

    \end{subfigure}
        \begin{subfigure}{0.33\textwidth}
        \includegraphics[width=\linewidth]{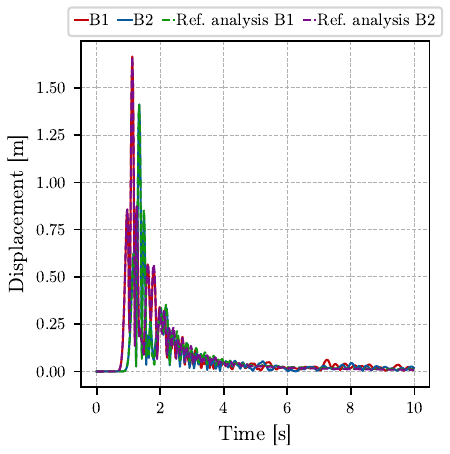}

    \end{subfigure}
    \\
    \begin{subfigure}{0.32\textwidth}
        \includegraphics[width=\linewidth]{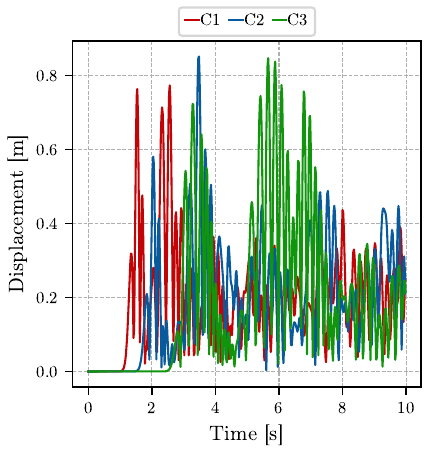}
        \caption{Standard implicit MPM}
    \end{subfigure}
    \begin{subfigure}{0.32\textwidth}
        \includegraphics[width=\linewidth]{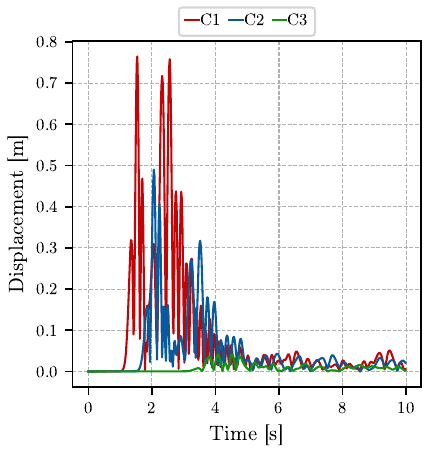}
        \caption{Proposed method}
    \end{subfigure}
        \begin{subfigure}{0.32\textwidth}
        \includegraphics[width=\linewidth]{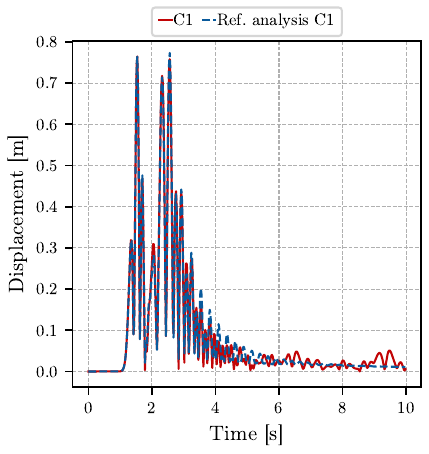}
        \caption{Proposed method with reference}
    \end{subfigure}
    \caption{2D elastic soil under point loading -- Test 3: displacement magnitudes comparisons between the standard implicit MPM scheme, the proposed method, and the reference simulation.}
    \label{fig:test3 graph}
\end{figure}

\begin{figure}[h!]
    \centering
    \includegraphics[width=\linewidth]{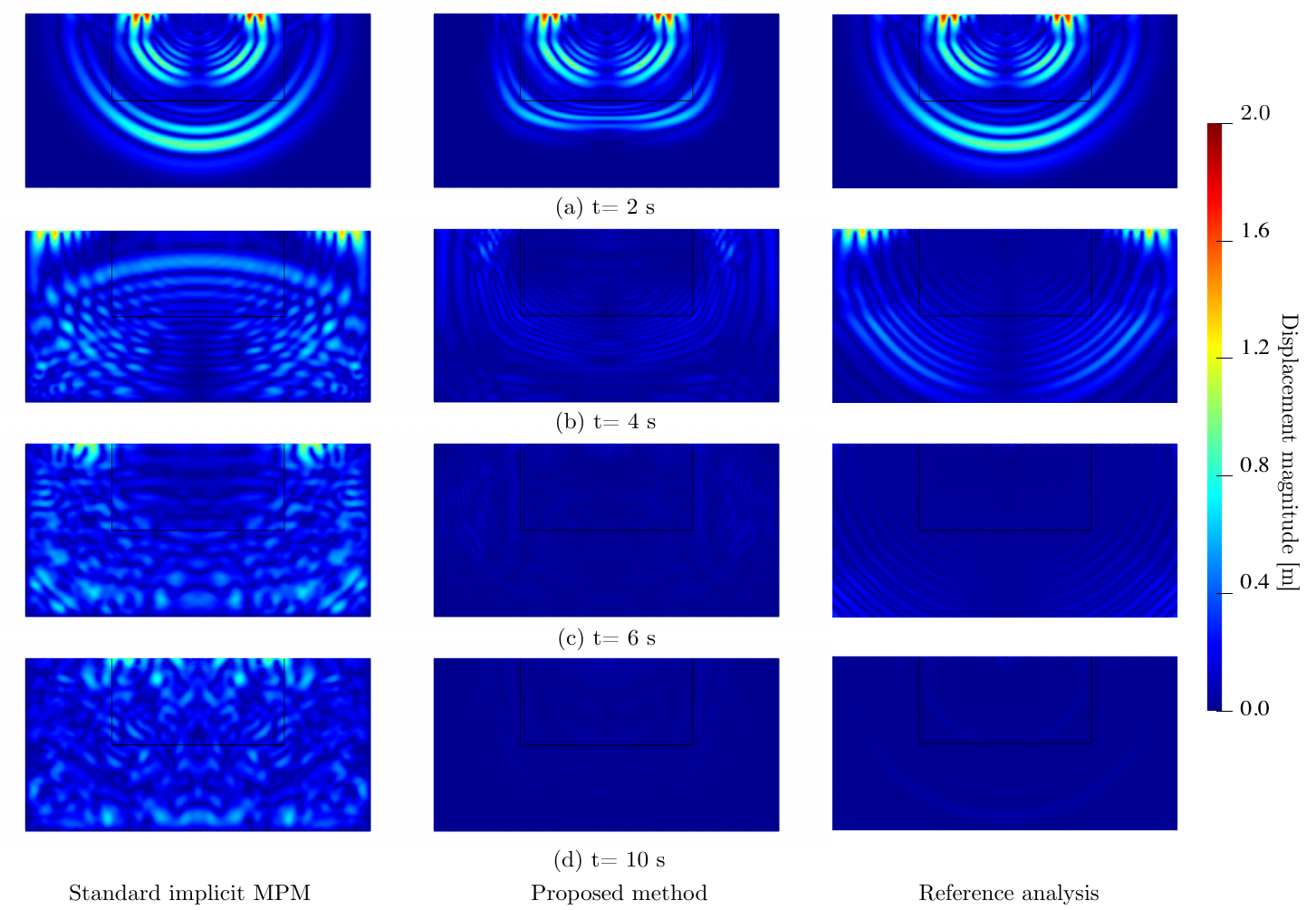}    
    \caption{2D elastic soil under point loading -- Test 3: displacement magnitude contours obtained by (left) standard implicit MPM, (center) the proposed method, and (right) the reference analysis with an enlarged domain at $t$ = 2.0, 4.0, 6.0, and 10.0 s.}
    \label{fig:test3 screenshot}
\end{figure}

Fig.~\ref{fig:test3 graph} presents the displacement magnitudes for Test 3. Due to the influence of multiple frequency loading \revision{given by the Ricker wavelet}, the results from the standard implicit MPM show larger magnitudes of reflected waves compared to the previous two test cases. This can be observed at receiver point A4, with notably higher amplitudes near the lateral boundary (upper right corner). After 5 seconds, the average ratio of maximum amplitude averaged about \revision{10.1\%} for A1-A4, \revision{13.5\%} for B1-B4, and \revision{11.6\%} for C1-C3. The damping performance of the method is further solidified by the evolution of displacement magnitude contours presented by Fig.~\ref{fig:test3 screenshot}. This demonstrates that the proposed method is adept not only at handling waves of singular frequencies but also complex waveforms that span multiple frequencies. Furthermore, by comparing the results at $t>4$ s, it is evident that body and surface waves, characterized by relatively large displacements, are significantly attenuated.

\revision{Despite the relatively good performance of the proposed PML method, it is important to highlight that the classical PML formulation, including the one we propose, still faces challenges in attenuating low-frequency surface waves. This is evident in Figs.~\ref{fig:test2 screenshot} and \ref{fig:test3 screenshot} at $t=4$ s, where the PML domain does not sufficiently absorb propagating waves near the free surface. Specifically, the proposed formulation (cf.~Eq.~\eqref{eq:dumping after inverse}) struggles to damp pure or quasi-one-dimensional motion, where compression or shear waves propagate exclusively in one direction in regions where the damping function $C_j$ is only applied in one direction, either $x$ or $y$ direction in 2D. This scenario is analogous to the conditions near the free surface observed here. Furthermore, these surface waves may get refracted back into the simulation domain, causing errors, as shown in Figs.~\ref{fig:test1 graph}, \ref{fig:test2 graph}, and \ref{fig:test3 graph}. While this error can be minimized by increasing the PML length ($L$), as demonstrated in \ref{app:changing_L} and suggested by \citet{festa2005interaction}, further extensions to the current formulation are necessary, e.g.~to include cut-off frequency \cite{festa2005interaction}.} 

\revision{Moreover, the current PML formulation is designed only to dampen propagating waves in the simulation domain, not evanescent waves. This approach was adopted to simplify the formulation, focusing on first establishing the PML framework within MPM. However, since evanescent waves can transform into body waves and generate spurious oscillations in the main domain, improving the PML formulation to include a damping term for both propagating and evanescent waves is necessary to enhance accuracy \cite{basu2003perfectly, basu2004perfectly, basu2009}.}

\subsection{Deformation of elasto-plastic embankment due to underground vibration}\label{elasticplastic_slope}
\begin{figure}[h!]
    \centering
    \includegraphics[width=0.8\linewidth]{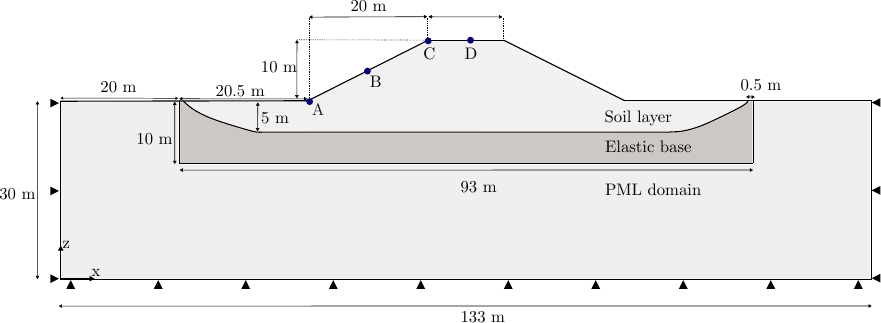}
    \caption{Symmetric elasto-plastic embankment under underground vibration: model geometry and boundary conditions.}
    \label{fig:embankment_model}
\end{figure}

In this section, we present a case study demonstrating the application of the proposed method to analyze embankment failure under dynamic loading. In this example, an elasto-plastic embankment is considered and modeled using the strain-softening Mohr-Coulomb model. Fig.~\ref{fig:embankment_model} illustrates the initial configuration of the model and the location of receiver points for the analysis; they are denoted as points A-D. The locations for the receiver points are as follows: \revision{A(38,30), B(46,34), C(57,39), D(69,39)}, all measured from the origin at the bottom left corner. The embankment slope is set at an angle of 26.57$^\circ$. The geometric configuration and material properties are derived from a prior investigation conducted by \citet{bui2015}, which explored the application of the SPH method in slope stability analysis. However, the geometry is slightly modified to ensure symmetry on both sides. 

Due to the absorbing mechanism of the PML formulation, it is important to note that the PML domain cannot directly interact with the elasto-plastic model\footnote{The PML theory works by introducing a coordinate transformation that attenuates the amplitude of elastic waves as they propagate into the PML domain. This transformation relies on the fact that the governing equations of motion for elastic waves (i.e.~Eq.~\eqref{eq:wave_equation3d}) are linear and can be manipulated mathematically to introduce damping without reflections.}. To address this limitation, our approach involves encasing the target domain within an elastic base, which is then surrounded by absorbing particles to simulate infinite boundary conditions effectively. To prevent unwanted stress concentration and plastic localization in sharp corners, a gradual transition between an elastic base and elasto-plastic soil is achieved using a quadratic function, as depicted in Fig.~\ref{fig:embankment_model} \revision{(more on this will be discussed in \Cref{rem:interface_geometries})}.

\begin{figure}[h!]
    \centering
    \includegraphics[width=0.35\linewidth]{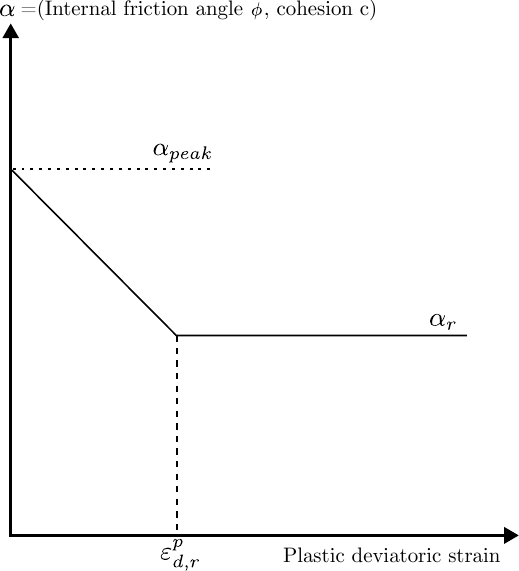}
    \caption{Illustration of the linear-strain-softening Mohr-Coulomb model behavior.}
    \label{fig:strain softening}
\end{figure}

The simulation initiates with static loading under gravity to establish the geo-static stress field, a crucial step before the dynamic analysis of confinement-dependent elasto-plastic materials. The specific parameters of the linear-strain-softening Mohr-Coulomb model are listed in \Cref{tab:parameter}. The soil layer parameters are determined based on \citet{bui2015} where they suggested that slope failure occurs under static loading when the peak friction angle and peak cohesion are $14^\circ$ and 6.9 kN/m, respectively. In the current model, we used these values as the residual values. For peak strength, we adopted the value at which static loading causes almost no displacement. The linear-strain-softening behavior is shown in Fig.~\ref{fig:strain softening}. Furthermore, the material parameters of the elastic base are set based on the elastic parameters of the soil layer, where the parameters of the PML domain follow. The length of the PML domain, $L$, is set to \revision{20 m, which corresponds to 40 PML elements}. Here, $\alpha_j$, $\beta$, and the fractional order of the visco-elastic model were fixed at 4, 1, and 0.95, respectively. The relaxed Young's modulus $E_0$ is set to 0.99 times Young's modulus of the absorbing particles, whereas the relaxation time $\tau$ was set to twice $\Delta t$. \revision{For a more comprehensive discussion on the rationale behind these parameter choices and their impact on the simulation results, please refer to the \ref{app:pml_param_sens_anal}.}

\begin{figure}[h!]
    \centering
    \includegraphics[width=0.5\linewidth]{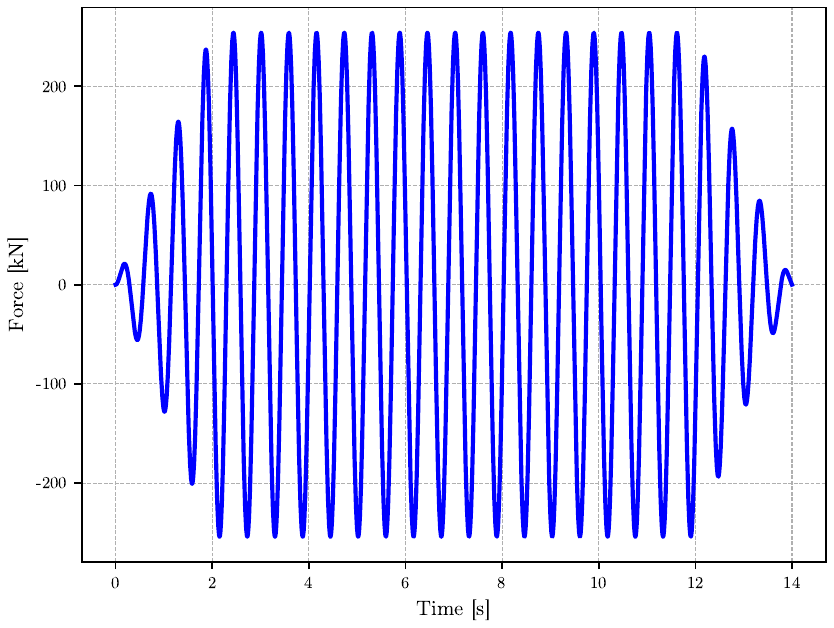}
    \caption{Symmetric elasto-plastic embankment under underground vibration: example of input wave over time for Test 1 with 254 N amplitude.}
    \label{fig:Input wave embankment}
\end{figure}

As depicted in Fig.~\ref{fig:Input wave embankment}, the seismic input for the analysis involves varying the amplitude of input between 254 N, 381 N, 508 N, 635 N, and 762 N to assess the model's response under different seismic intensities. The inputs shown in Fig.~\ref{fig:Input wave embankment} are applied to the particles in the elastic base as seismic body forces. Although such input methods are not generally used for seismic loading, which commonly employs moving traction BC at the finite element nodes, this approach was adopted to avoid handling nonconforming Neumann traction BC, which is known to be cumbersome in MPM; see \cite{bing2019b, NAnda2021}. In the MPM simulations, we use a quadrilateral background grid with a cell size of $h = 0.5$ m, resulting in a total of \revision{23,940 cells}. The domain is entirely discretized into \revision{68,960 particles} (20,000 material points and \revision{48,960 absorbing particles}), initially arranged in a configuration of $2\times 2$ particles per cell. The time step is set as $\Delta t = 0.001$ s. \revision{As the elasto-plastic analysis requires considerably more time to perform, reference simulation with an enlarged domain is not performed for this test.}

\begin{table}[h!]
  \centering
  \caption{Symmetric elasto-plastic embankment under underground vibration: material parameters for different layers.}
  \label{tab:parameter}
  \begin{tabular}{||l|c||}
    \hline
    Parameter & Value \\
    \hline    \hline
    $\mathbf{Elastic\ base}$ & \\
    Young’s modulus, \( E_{base} \) (MPa) & 100 \\
    Poisson’s ratio, \( \nu_{base} \) & 0.33 \\
    Density, \( \rho_{base} \) (kg/m\(^3\)) & 2041 \\
    \hline
    \( \mathbf{Soil\ layer}\) & \\
    Young’s modulus, \( E_{layer} \) (MPa) &  100 \\
    Poisson’s ratio, \( \nu_{layer} \) & 0.33 \\
    Density, \( \rho_{layer} \) (kg/m\(^3\)) & 2041 \\
    Peak internal friction angle, $\phi_{peak}$ ($^\circ$)  &16\\
    Residual internal friction angle, $\phi_r$ ($^\circ$)  &14\\
    Peak cohesion, $c_{peak}$  (kN/m) &8\\
    Residual cohesion, $c_r$  (kN/m) &6.9\\
    Residual plastic deviatoric strain, $\varepsilon_{d,r}^{p}$  & 0.13               \\
    Dilatancy angle \( \psi \) ($^\circ$) & 9\\
    \hline
    \( \mathbf{PML\ domain}\) & \\
    Young’s modulus, \( E_{base} \) (MPa) & 100 \\
    Poisson’s ratio, \( \nu_{base} \) & 0.33 \\
    Density, \( \rho_{base} \) (kg/m\(^3\)) & 2041 \\
    Maximum damping ratio, \(\alpha_j\)  & 4\\
    Damping power, $\beta$ & 1\\
    Visco elastic relaxed youngs modulus, \(E_0\) (MPa) & 99 \\
    Visco elastic fractional order, \(\alpha\)  &   0.95\\
    Visco elastic relaxation time, \(\tau\) (s)      &   0.002\\
    Rayleigh damping factor, \(\alpha_M \)   & 1.0\\
    \hline
  \end{tabular}
\end{table}

\begin{figure}[h!]
    \centering
    \includegraphics[width=0.95\linewidth]{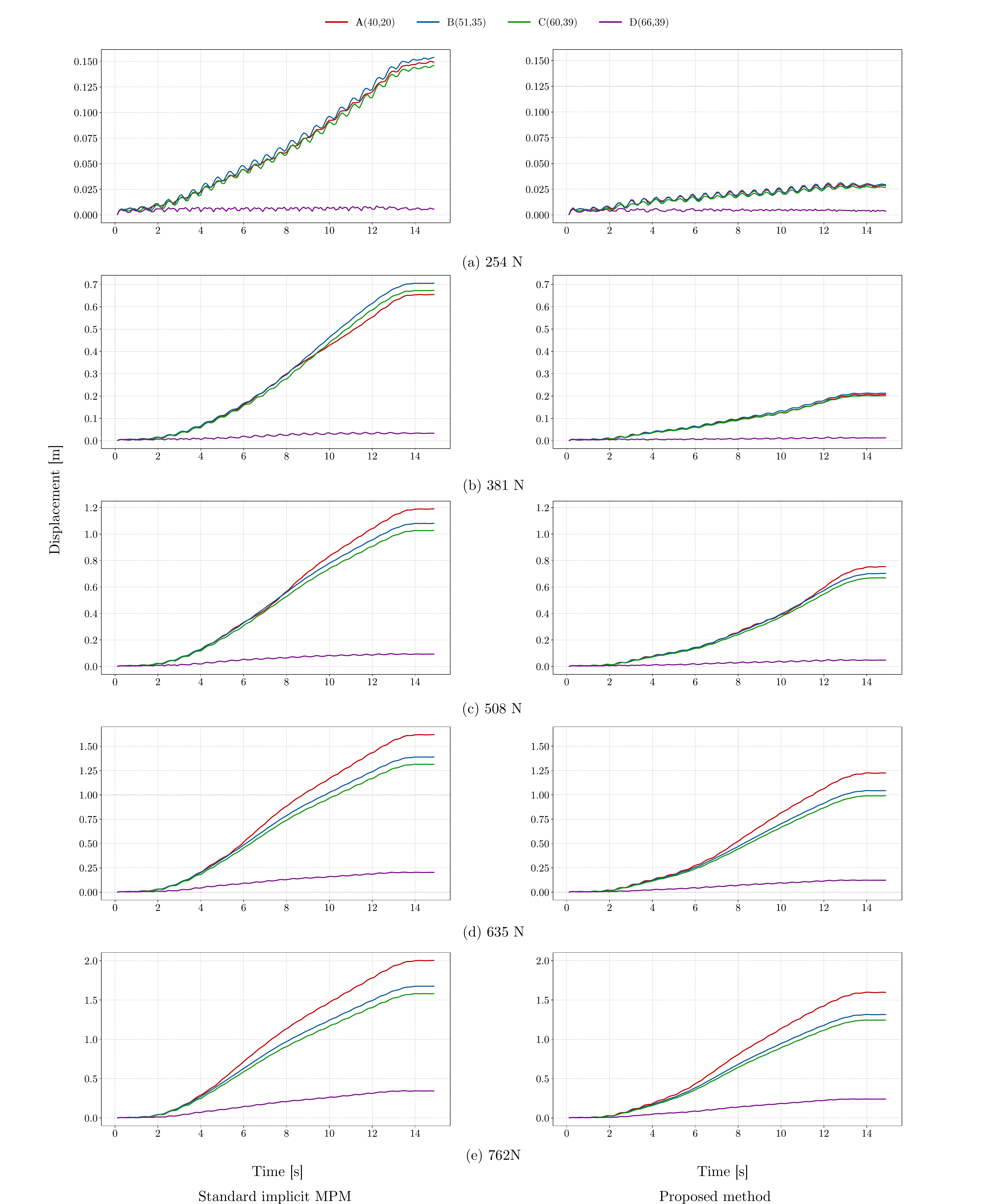}
    \caption{Symmetric elasto-plastic embankment under underground vibration: displacement magnitudes measured at different measurement points at five different loading cases.}
    \label{fig:slope_graph}
\end{figure}

\begin{figure}[h!]
\centering
  \begin{minipage}[b]{0.7\linewidth}
    \centering
    \includegraphics[width=\columnwidth]{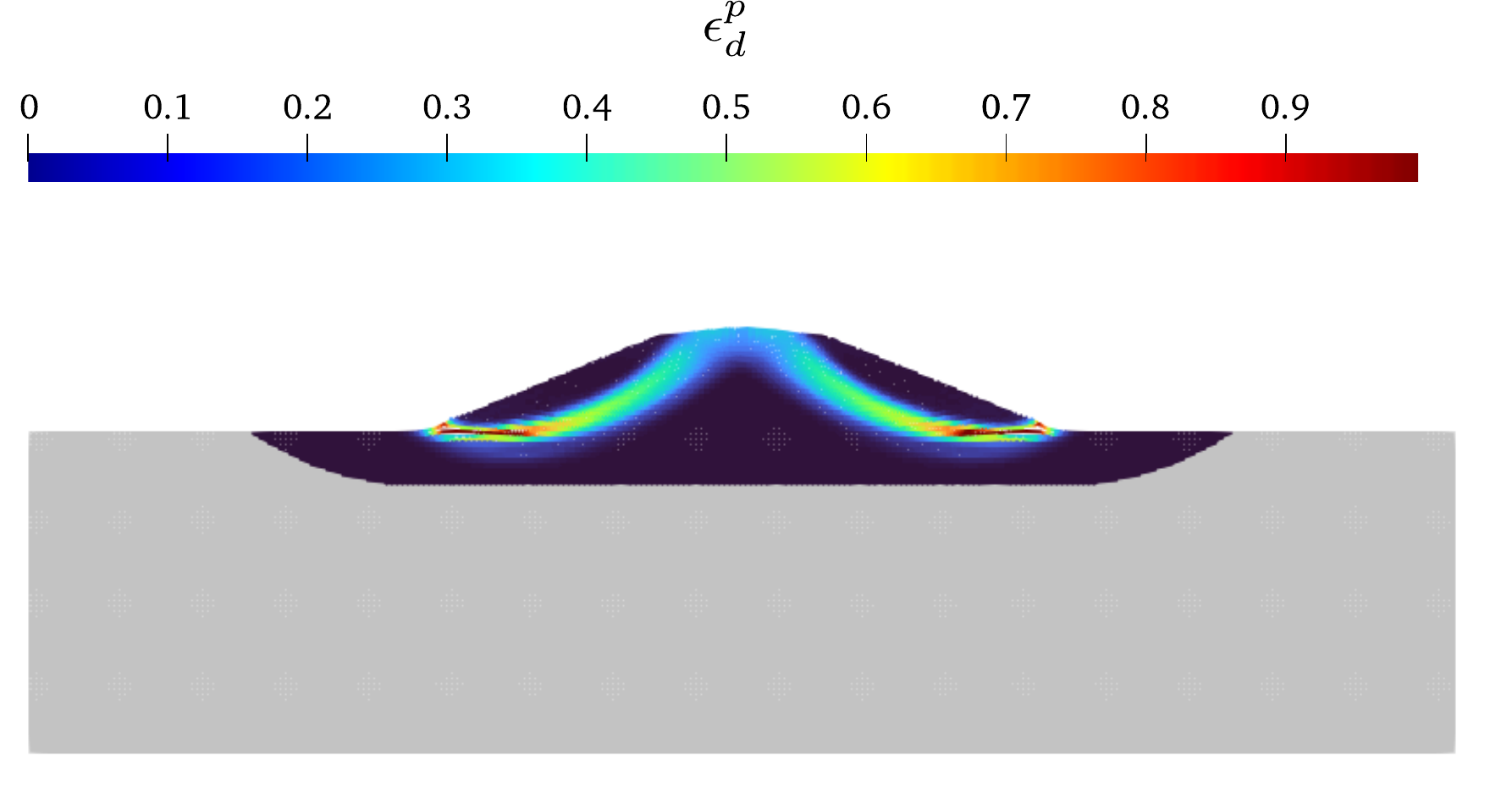}
    \subcaption{Standard implicit MPM}
  \end{minipage}
  \begin{minipage}[b]{0.7\linewidth}
    \centering
    \includegraphics[width=\columnwidth]{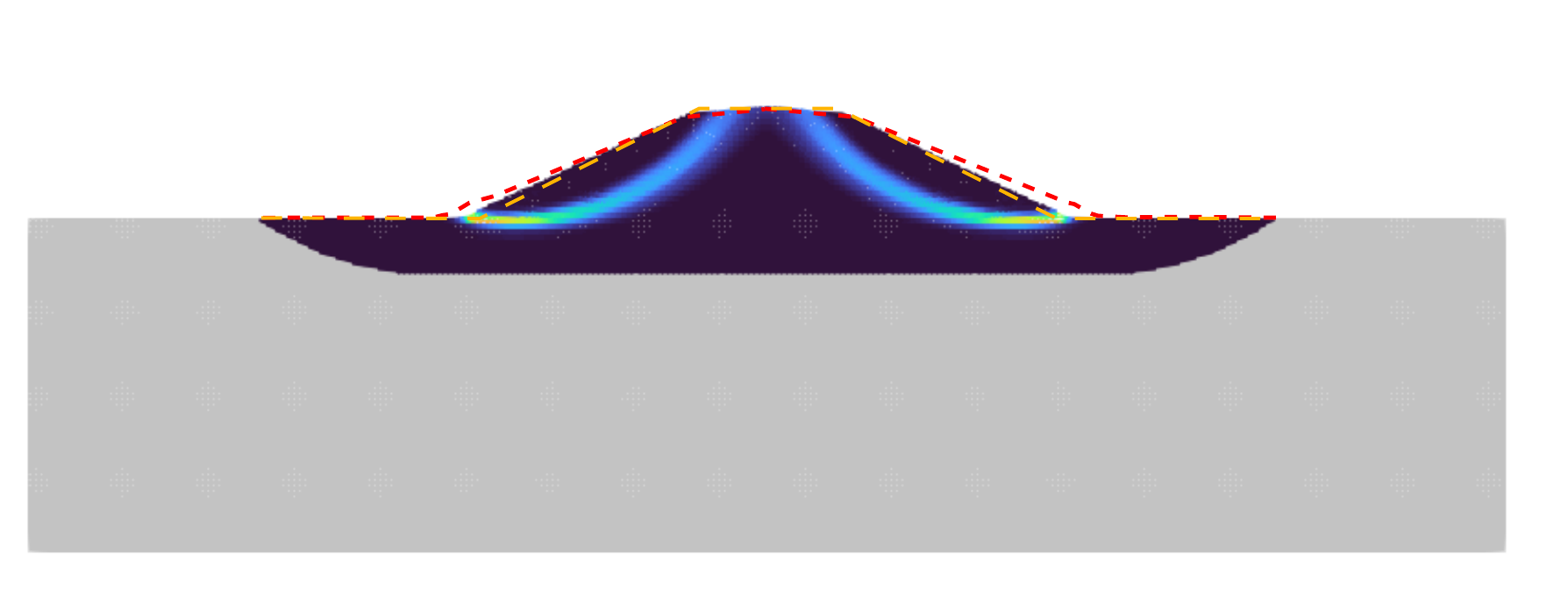}
    \subcaption{Proposed method}
  \end{minipage}
  \caption{Symmetric elasto-plastic embankment under underground vibration: comparison of the plastic strain of standard implicit MPM (top) and the proposed method (bottom) at $t=15$ s for Test 5. The red dashed line plots the deformed surface of the standard implicit MPM results for comparison and the orange dashed line plots the original surface of the embankment.}
  \label{fig:compare_strain}
\end{figure}

Fig.~\ref{fig:slope_graph} plots the displacement magnitudes obtained from the proposed method compared to the results of standard implicit MPM performed without the absorbing particles. Here, the dynamic loading conditions range from the amplitude of 254 N to 762 N, which are ordered in increasing order from top to bottom. Here, we refer to them as Test 1-5, respectively. From Fig.~\ref{fig:slope_graph} (a)-(e), we can observe that the proposed method induced a comparatively smaller displacement magnitude compared to the standard implicit MPM results. This indicates that by attenuating outgoing waves via absorbing particles, we can reduce the overall energy in the system which, if it is not absorbed properly, may induce excessive estimation of soil deformation.

Fig.~\ref{fig:compare_strain} shows the plastic strain distribution of the standard implicit MPM and the proposed method for Test 5, with the red dashed line representing the deformation shape of the standard implicit MPM in Fig.~\ref{fig:compare_strain} (b). In comparison to the standard implicit MPM results, the proposed method has a narrower strain zone, indicating that the strain is not distributed over the entire embankment. The shape of the deformed slopes also differs from that of the dashed line because the displacement induced by the refraction waves can be optimally suppressed.

\begin{figure}[h!]
    \centering
    \includegraphics[width=0.5\linewidth]{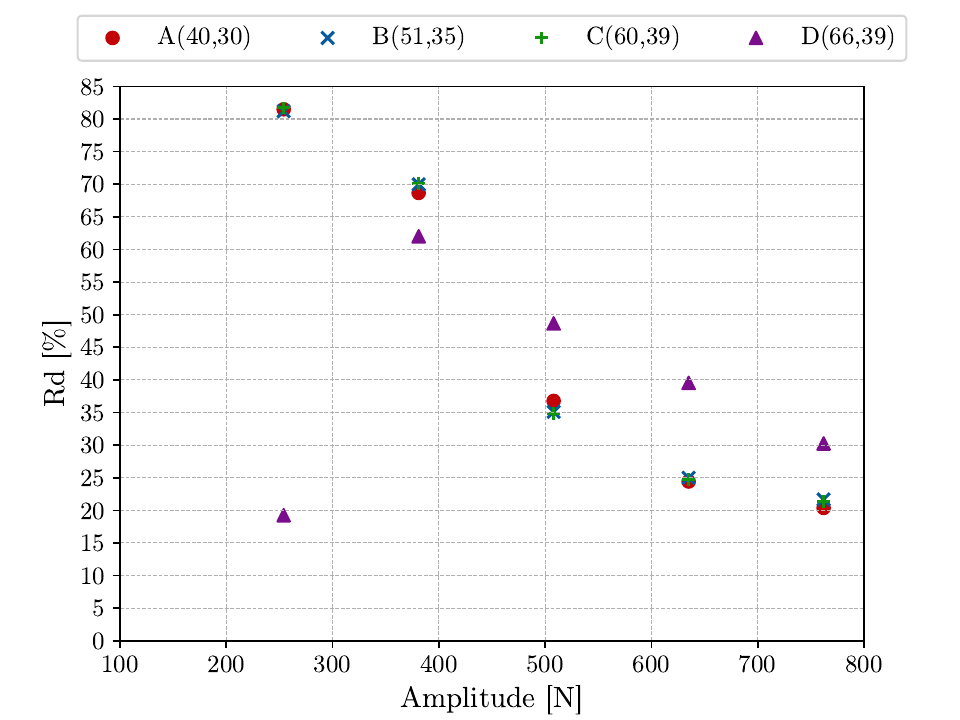}
    \caption{Symmetric elasto-plastic embankment under underground vibration: relative displacement change ratio obtained at receiver points for different vibration amplitudes.}
    \label{fig:displacement ratio}
\end{figure}

To measure the damping performance quantitatively, we define the relative displacement change ratio $R_d$ as follows:
\begin{equation}
    R_d =  \frac{u_{\text{{Im}}} - u_{\text{Pr}}}{u_{\text{{Im}}}}\times 100 \%\,,
\end{equation}
where $u_{\text{{Im}}}$ and $u_{\text{Pr}}$ denote the displacement magnitudes of the standard implicit MPM and the proposed method, respectively. The obtained relative displacement change ratio is plotted in Fig.~\ref{fig:displacement ratio} for all five cases. \revision{First, notice that in all tests, the values of $R_d$ for points A, B, and C are approximately the same, while point D shows a slightly different $R_d$ value compared to the other points. This is because the displacements of points A-C consist of more dominated horizontal deformation and vertical deformation, whereas point D moves only vertically due to the symmetric geometry. In Test 1 (254 N), points A-C exhibited relative displacement change ratios of approximately 80\%. As the input amplitude increases, the relative displacement change ratio shows a decreasing trend. This indicates that the displacement caused by the input wave itself becomes more dominant as the amplitude increases, inducing larger deformation relative to the deformation caused by the undamped outgoing waves.} 

\begin{remark}
\label{rem:interface_geometries}
     \revision{We tested the effect of interface geometry between elastic-base and elasto-plastic soils, as illustrated in Fig.~\ref{fig:compare_shape_0002i}. This analysis considers Test 5, which involves the highest amplitude of base shaking, i.e.~762 N. The figure shows the distribution of plastic strain for different interface geometries, where the shapes are modified by varying the coefficients of a quadratic function, i.e.~from left to right: ${5}/{15^2}$, ${5}/{7.5^2}$, and 0 (a jump). We denote these cases as "long," "medium," and "sharp" interfaces, respectively. Here, the lengths of the interfaces differ and are determined so that all intersect at the height of 2.5 m (half of the elasto-plastic soil thickness).} 
     
     \revision{As shown in the figures, concentrated plastic strains develop at the surface during shaking. In the medium and sharp cases, the plastic strain concentrations are more than twice as large as those in the long interface case (as shown in Fig.~\ref{fig:compare_shape_0002i}, with the color bar presented on a logarithmic scale for better visualization). While this effect is not particularly pronounced in this example, the development of plastic regions could induce inaccuracies in surface wave propagation. In certain cases, especially with more sensitive soils, plastic strain concentration may also lead to shear strength softening, resulting in excessive plastic deformation near the interface. To mitigate this issue, a more gradual transition between the elastic base and elasto-plastic soils is recommended for improved stability and accuracy.}
\begin{figure}[h!]
    \centering
    \includegraphics[width=\linewidth]{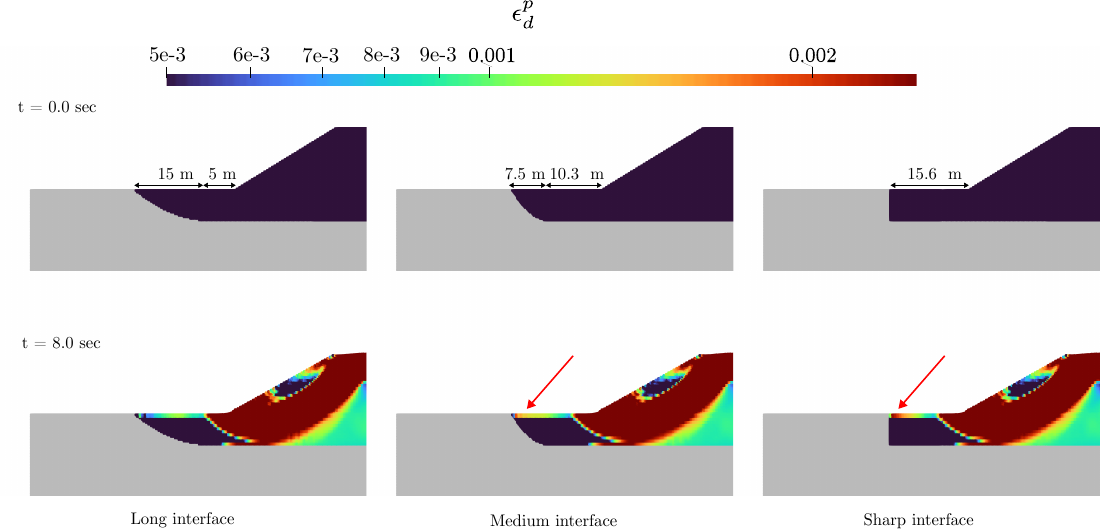}
    \caption{Symmetric elasto-plastic embankment under underground vibration: plastic deviatoric strain comparison for different interface shape between elasto-plastic soil layer and elastic base at $t=8$ s. The geometry of the interface is modified by varying the coefficients of a quadratic function (from left to right: ${5}/{15^2}$, ${5}/{7.5^2}$, and 0). Red arrows highlight the plastic strain concentration at the free surface. Here, a log-scale color contour is considered for better visualization of the magnitude range.}
    \label{fig:compare_shape_0002i}
\end{figure}
\end{remark}

\subsection{Earthquake-induced slope failure analysis}\label{slope_failure}
\begin{figure}[h!]
    \centering
    \includegraphics[width=\linewidth]{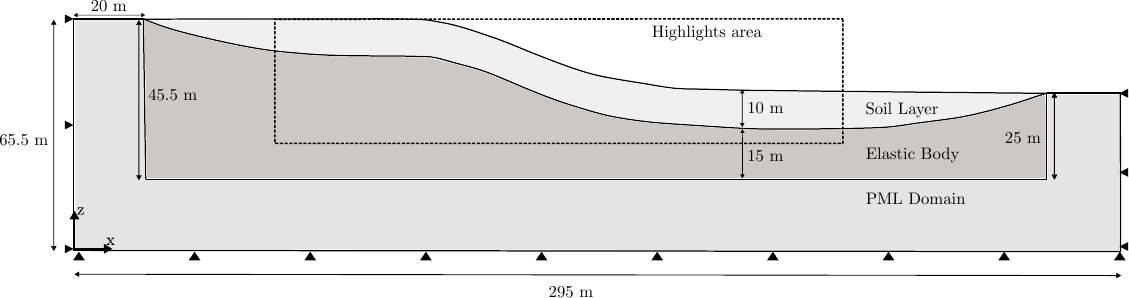}
    \caption{Earthquake-induced slope failure analysis: model geometry and boundary condition.}
    \label{fig:gaussian model}
\end{figure}

\begin{figure}[h!]
    \centering
    \includegraphics[width=0.6\linewidth]{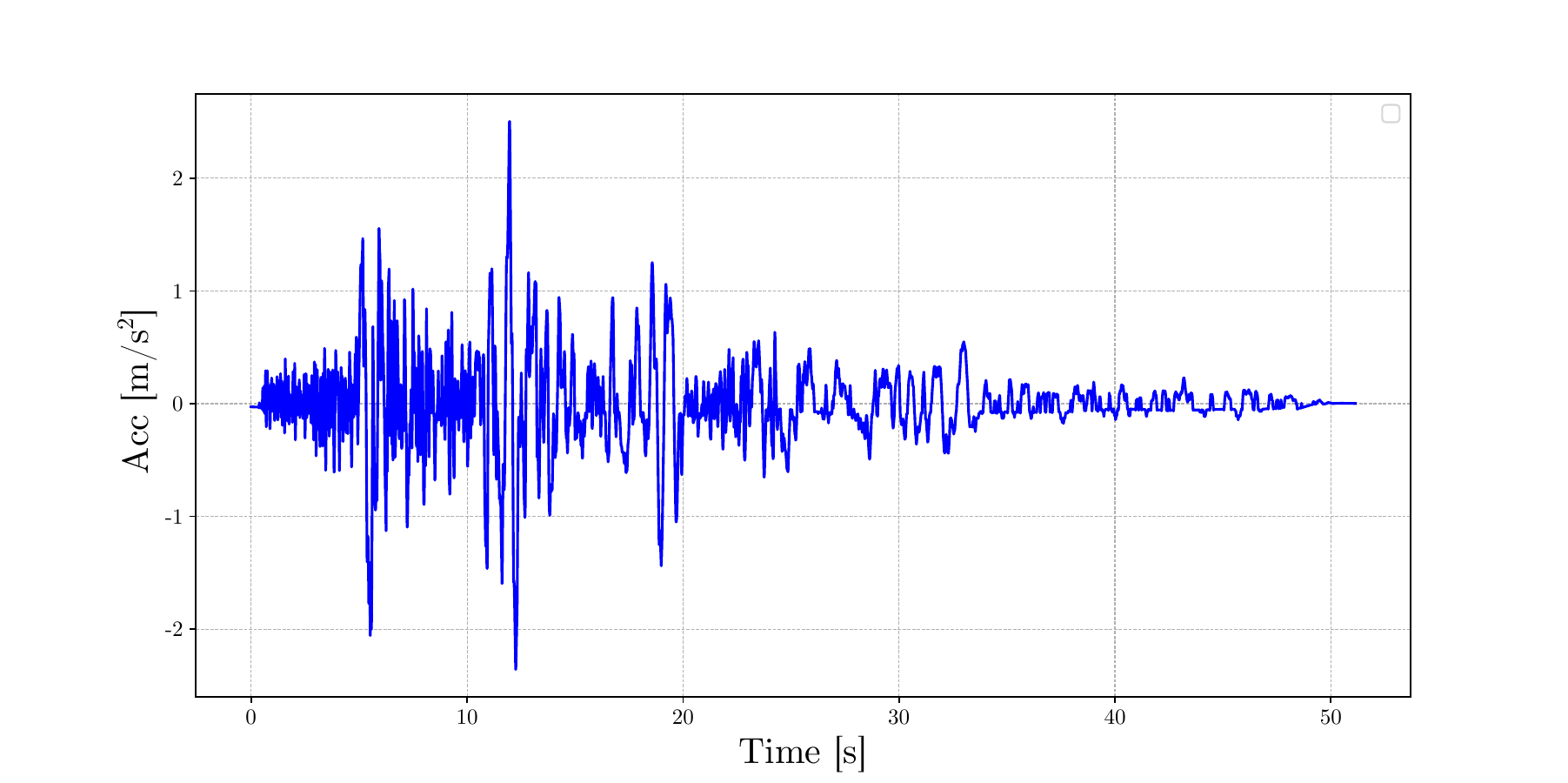}
    \caption{Earthquake-induced slope failure analysis: seismic input wave (the Imperial Valley event in 1979 (RSN 165, H2 direction \cite{ancheta})).}
    \label{fig:acc plt}
\end{figure}

An example case of asymmetry base shaking with an elasto-plastic slope is conducted in this section following the numerical example done by \citet{kohler}. The geometry settings are defined in Fig.~\ref{fig:gaussian model}. Here, we modify the geometry to be enclosed in an elastic base. Input waves recorded from the Imperial Valley event in 1979 (RSN 165, H2 direction) were used following \citet{kohler}. Waveforms were obtained from the PEER (Pacific Earthquake Engineering Research) strong-motion database \cite{ancheta}, and are shown in Fig.~\ref{fig:acc plt}.

The MPM simulations are performed using quadrilateral background grids with a cell size of $h = 0.5$ m, resulting in a total of \revision{94,400 cells}. The entire domain is discretized as \revision{254,619 particles} (137,899 material points and \revision{116,720 absorbing particles}) which are initially arranged in 2$\times$2 particles per cell configuration. The initial volume of material points is assumed to be uniform. Moreover, the time step is set as $\Delta t$ = 0.001 s. \revision{Similar to \Cref{elasticplastic_slope}, reference simulation with an enlarged simulation domain is not performed for this numerical example.} In addition, we utilized the B-bar method \cite{hughes1977equivalence, chandra2023high} to mitigate volumetric-locking effects associated with fast undrained deformation. The considered material parameters and parameters for absorbing particles are listed in \Cref{tab:parameter2} as adopted from \citet{kohler}. \citet{kohler} used the von Mises yield criterion, whereas we used the Tresca model as it aligns better with the shear-dominated failure mechanisms of undrained cohesive soils. Even though the considered yield criteria are different, the linear strain softening model is considered to be the same, with residual undrained shear strength, $Su_r$, and the residual plastic deviatoric strain, $\varepsilon_{d,r}^{p}$, are set to be the same values as \citet{kohler}. The behavior of the linear strain softening model is similar to that shown in Fig.~\ref{fig:strain softening}. The parameters of the absorbing particles are set in the same way as described in \Cref{elasticplastic_slope}. The length of the PML domain, $L$, is set to \revision{20 m, which corresponds to 40 PML elements. This setting is based on the sensitivity study elaborated in \ref{app:pml_param_sens_anal}.}


\begin{table}[h!]
  \centering
  \caption{Earthquake-induced slope failure analysis: material and PML parameters.}
  \label{tab:parameter2}
  \begin{tabular}{||l|c||}
    \hline
    Parameter & Value \\
    \hline
    \hline
    $\mathbf{Elastic\ base}$ & \\
    Young’s modulus, \( E_{base} \) (MPa) & 250 \\
    Poisson’s ratio, \( \nu_{base} \) & 0.25 \\
    Density, \( \rho_{base} \) (kg/m\(^3\)) & 2200 \\
    \hline
    \( \mathbf{Soil\ layer}\) & \\
    Young’s modulus, \( E_{layer} \) (MPa) &  40\\
    Poisson’s ratio, \( \nu_{layer} \) & 0.495 \\
    Density, \( \rho_{layer} \) (kg/m\(^3\)) & 2200 \\
    Peak undrained strength, $Su_p$  (kN/m) &35\\
    Residual undrained strength, $Su_r$  (kN/m) &19.44\\
    Residual plastic deviatoric strain, $\varepsilon_{d,r}^{p}$:  & 0.231               \\
    \hline
    \( \mathbf{PML\ domain}\) & \\
    Young’s modulus, \( E_{base} \) (MPa) & 250 \\
    Poisson’s ratio, \( \nu_{base} \) & 0.25 \\
    Density, \( \rho_{base} \) (kg/m\(^3\)) & 2200 \\
    Maximum damping ratio, \(\alpha_j\)  & 4\\
    Damping power, \(\beta\) & 1\\
    Visco elastic relaxed youngs modulus, \(E_0\) (MPa) & 247.5 \\
    Visco elastic fractional order, \(\alpha\)  &   0.95\\
    Visco elastic relaxation time, \(\tau\) (s)      &   0.002\\
    Rayleigh damping factor, \(\alpha_M \)   & 1.0\\
    \hline
  \end{tabular}
\end{table}

\begin{figure}[h!]
    \centering
    \includegraphics[width=\linewidth]{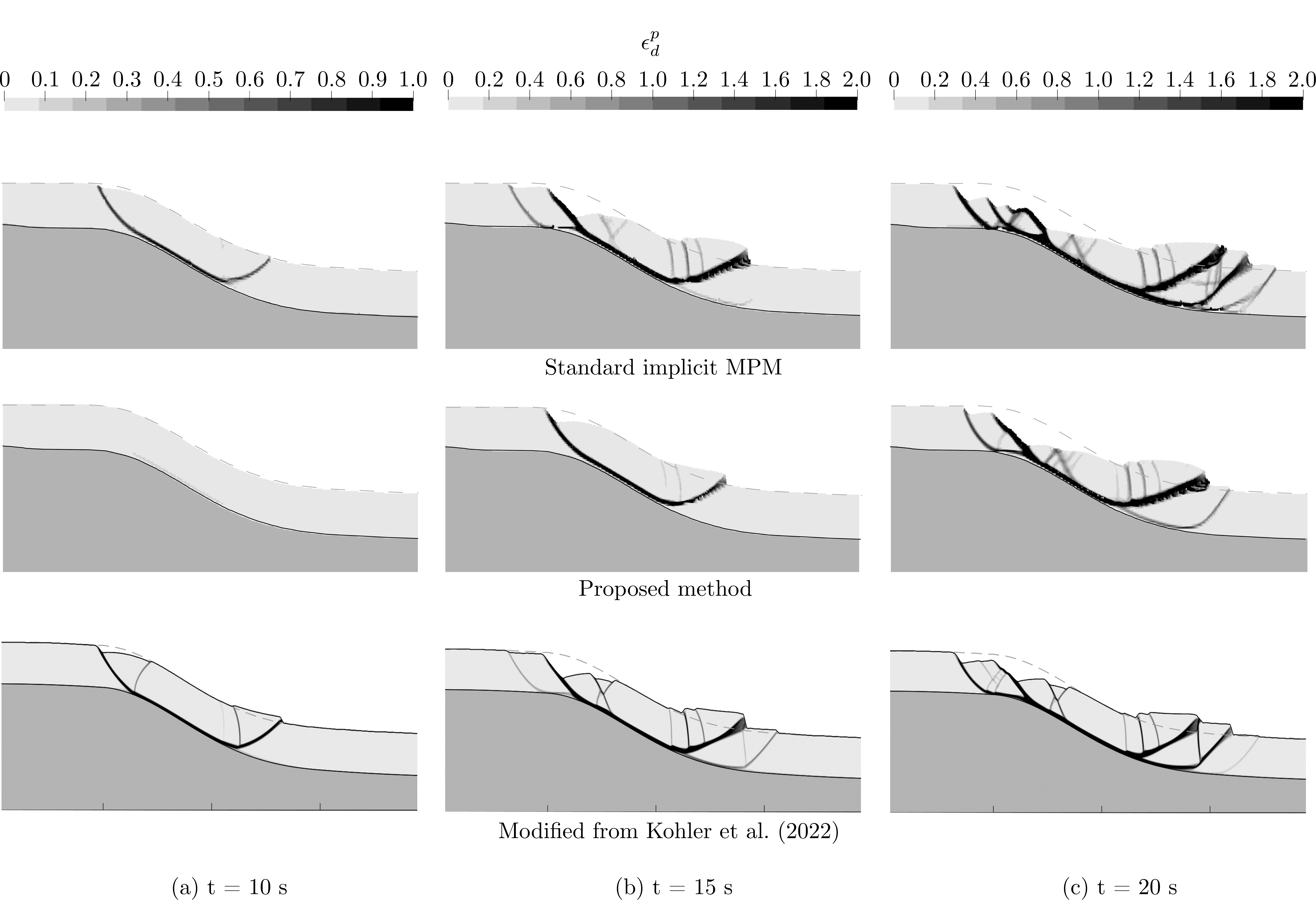}
    \caption{Earthquake-induced slope failure analysis: comparison of the evolution of plastic strain from the standard implicit MPM (top), the proposed method (middle) and numerical results of \citet{kohler} (bottom) at $t =$ 10.0 s, 15.0 s, and 20.0 s.}
    \label{fig:result gaussian}
\end{figure}

The obtained numerical results are presented in Fig.~\ref{fig:result gaussian}, where the plastic strain distribution at different snapshots is plotted for comparison. The figure presents three rows of results: the top row features standard implicit MPM results without absorbing particles, the middle row shows the results of the proposed MPM-PML method, and the bottom row is taken from the analysis results performed by \citet{kohler}. The results without absorbing particles reveal that a slip surface forms at 10 sec and the slope collapses catastrophically due to strength softening over time. The onset and propagation of the shear bands are similar to the simulations performed by \citet{kohler}, despite differences in mesh size and analysis conditions. It is worth emphasizing that, \citet{kohler} considers a mesh size of $h=0.25$ m, which is half of the currently considered mesh size, with 3$\times$3 particle per cell, which is higher compared to our simulations. They also employed a viscous surface traction BC to absorb outgoing waves. However, the accuracy of such methods in MPM is known to present inaccuracy and challenges in large-deformation settings as the material boundary becomes nonconforming to the element boundary. On the other hand, in the results of the proposed method, a slip surface is first formed at around $t = 15$ s, and through comparisons at a later stage, the generated landslide is much less severe compared to the standard implicit MPM simulations and those performed by \citet{kohler}. This indicates that the addition of absorbing particles can effectively dampen the refracted seismic waves from over-exciting the slope failures, potentially enabling a more precise analysis.

\begin{figure}[h!]
    \centering
  \begin{minipage}[b]{0.45\textwidth}
    \includegraphics[width=\linewidth]{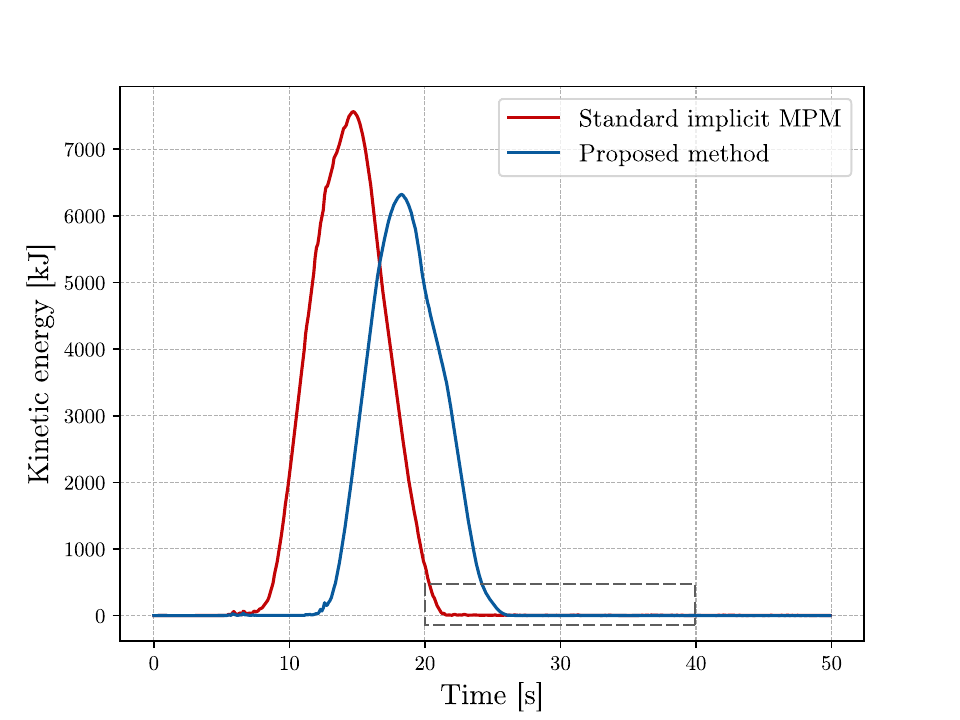}
    \subcaption{$t = 0\sim50 s$}
  \end{minipage}
  \begin{minipage}[b]{0.45\textwidth}
    \includegraphics[width=\textwidth]{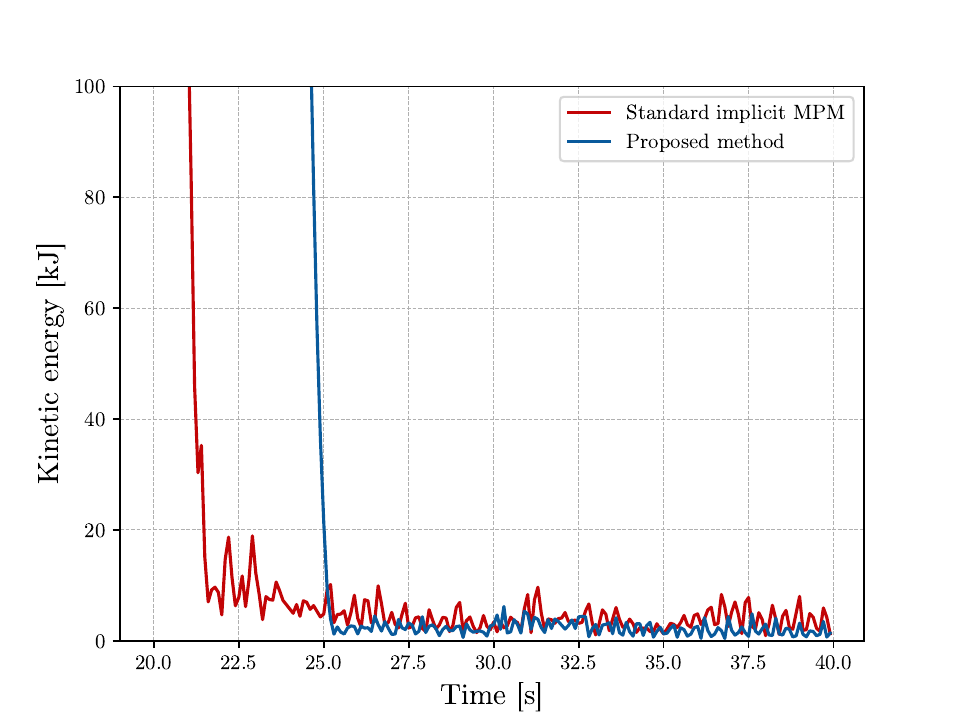}
    \subcaption{$t = 20\sim40 s$}
  \end{minipage}
  \caption{\revision{Earthquake-induced slope failure analysis: (a) comparison of the evolution of kinetic energy and (b) a zoomed-in region around $t=20\sim40$s. The zoomed plot reveals minor but non-vanishing fluctuations in kinetic energy due to the ongoing seismic excitation, with values significantly smaller than the peak increase in kinetic energy resulting from the transient post-failure slope deformation.}}
  \label{fig:kinetic_energy}
\end{figure}

To measure the evolution of soil deformation during earthquake shaking, we approximate the total kinetic energy of the system following:
\begin{equation}
    \Pi_K = \sum_{p=1}^{N_p}\frac{1}{2}m_p \widetilde{\tb v}_p \cdot \widetilde{\tb v}_p\,, \qquad \mathrm{where} \quad \widetilde{\tb v}_p = \sum_{I=1}^{n_n} N_{Ip} \tb v_I\,.
\end{equation}
The evolution of the kinetic energy $\Pi_K$ of the soil layer without the elastic base and the PML domain over time is plotted by Fig.~\ref{fig:kinetic_energy}. In the standard implicit MPM, the kinetic energy increases at a constant rate from the formation of the sliding surface (at around $t=6$ s) and then gradually decreases after the majority of the slopes have failed. The results of the proposed method show that there is a time lag of about 6 to 7 seconds before the kinetic energy starts to increase compared to the standard implicit MPM, and the peak value is also smaller than that of the standard implicit MPM result. Such continuous energy transitions during earthquake shaking can only be obtained by considering continuous simulations of earthquakes and slope failure. As presented in a recent study conducted by \citet{sordo4783551sequential}, this was difficult to be achieved using subsequent analysis of FEM and MPM. 

\section{Conclusions}
\label{sec:conclusion}
This study proposes the implementation of PML theory in the dynamic MPM framework utilizing a set of absorbing particles. The governing equations of PML in the time domain proposed by \citet{chen2022} are integrated into the implicit MPM along with the enhanced damping strategies. The discretized form of the modified balance equations is derived and proposed, which can be solved concurrently with the standard elasto-plastic implicit MPM equations employing the Newton-Raphson algorithm.

The effectiveness of the proposed approach is confirmed by simulating different types of numerical examples, considering impulse loading on elastic bodies and input waves on symmetric and asymmetric bases. The method has been shown to effectively dampen outgoing waves adequately and minimize wave reflections at boundaries. It has been verified to handle large deformations and complex wave patterns with remarkable accuracy. In particular, the analysis of elastoplastic slope cases highlighted the influence of absorbing particles on the timing and developed mechanism of slope failure, demonstrating that the generated simulation is comparatively more stable compared to conventional methods. This development enhances MPM's capabilities in geotechnical analysis and contributes to a better understanding of the complex mechanisms that govern earthquake-induced landslides. While the proposed method shows promising results, further validation tests with controlled experiments or recorded past events are necessary to confirm the accuracy of simulating dynamic, time-dependent phenomena.

\revision{Immediate work is needed to extend the accuracy of the proposed model by incorporating the coordinate stretching term, which will dampen not only propagating waves but also evanescent waves.} Future challenges include the need to consider new ways of handling seismic input waves that can handle nonconforming Neumann BC and, at the same time, avoid direct inputs of external forces into the PML domain. Furthermore, as the elasto-plastic model and the PML domain cannot be in direct contact, this study necessitates a more in-depth examination of how elastic bodies are managed around the elasto-plastic model. The application of the proposed method to more complex scenarios, including hydro-mechanical coupled conditions, will be a major research theme in the future, as will the analysis of large ground displacements, including the presence of liquefaction.

\section*{CRediT authorship contribution statement}
\textbf{J. Kurima}: Conceptualization, Methodology, Software, Validation, Visualization, Formal analysis, Writing - original draft. \textbf{B. Chandra:} Conceptualization, Methodology, Software, Formal analysis, Project administration, Writing - review and editing. \textbf{K. Soga:} Resources, Supervision, Writing - review and editing.

\section*{Acknowledgments}
This work is supported by the Japan Society for the Promotion of Science Kakenhi with grant numbers JP23K19135, JP24K17347, and JP21H01418. The authors acknowledge the support of the Donald H. McLaughlin Chair Fund of the University of California, Berkeley which partly funded this work. The authors thank Connor Geudeker from UC Berkeley for their help and insightful feedback in reviewing this manuscript.

\section*{Declaration of competing interest}
The authors declare that they have no conflict of interest.

\section*{Data availability statement}
The data that support the findings of this study are available from the corresponding authors upon reasonable request.

\appendix

\section{Sensitivity analysis of PML parameters for elastodynamics problems}
\label{app:pml_param_sens_anal}

\revision{This appendix provides useful guidelines for selecting the parameters required when using the proposed PML elements and absorbing particles. These parameters include mesh size, the length of the PML domain $L$, PML damping parameters $\alpha_j$ and $\beta$, as well as viscoelastic and Rayleigh damping parameters. To enable a comprehensive and in-depth investigation, we consider the 2D elastic soil problem under point-source impact loading described in \Cref{elastic_2d}. Specifically, we examine the accuracy of the measured displacements at receiver points under Test 2 loading (cf.~Fig.~\ref{fig:Input wave2}), within the time interval $t = 5\sim10$ s, and compare them to a reference simulation with an extended domain.}


\subsection{Changing mesh size, $h$}

\revision{In this section, we evaluate the damping effects of the proposed method with respect to mesh size. In the analysis conducted in \Cref{elastic_2d}, the cell size was set to 20 m, and the particle size to 10 m (4 particles per cell in 2D). Here, we performed analyses with mesh sizes of 10 m and 40 m, and corresponding particle sizes of 5 m and 20 m (keeping the 4 PPC ratio constant), respectively. We refer to the cases with mesh sizes of 40 m, 20 m, and 10 m as coarse, medium, and fine cases. All other parameters remained identical to those given in \Cref{elastic_2d}, in particular, we kept the PML distance $L$ to be constant as $L=1000$ m.}

\begin{figure}[h!]
    \centering
         \centering
         \includegraphics[width=0.55\textwidth]{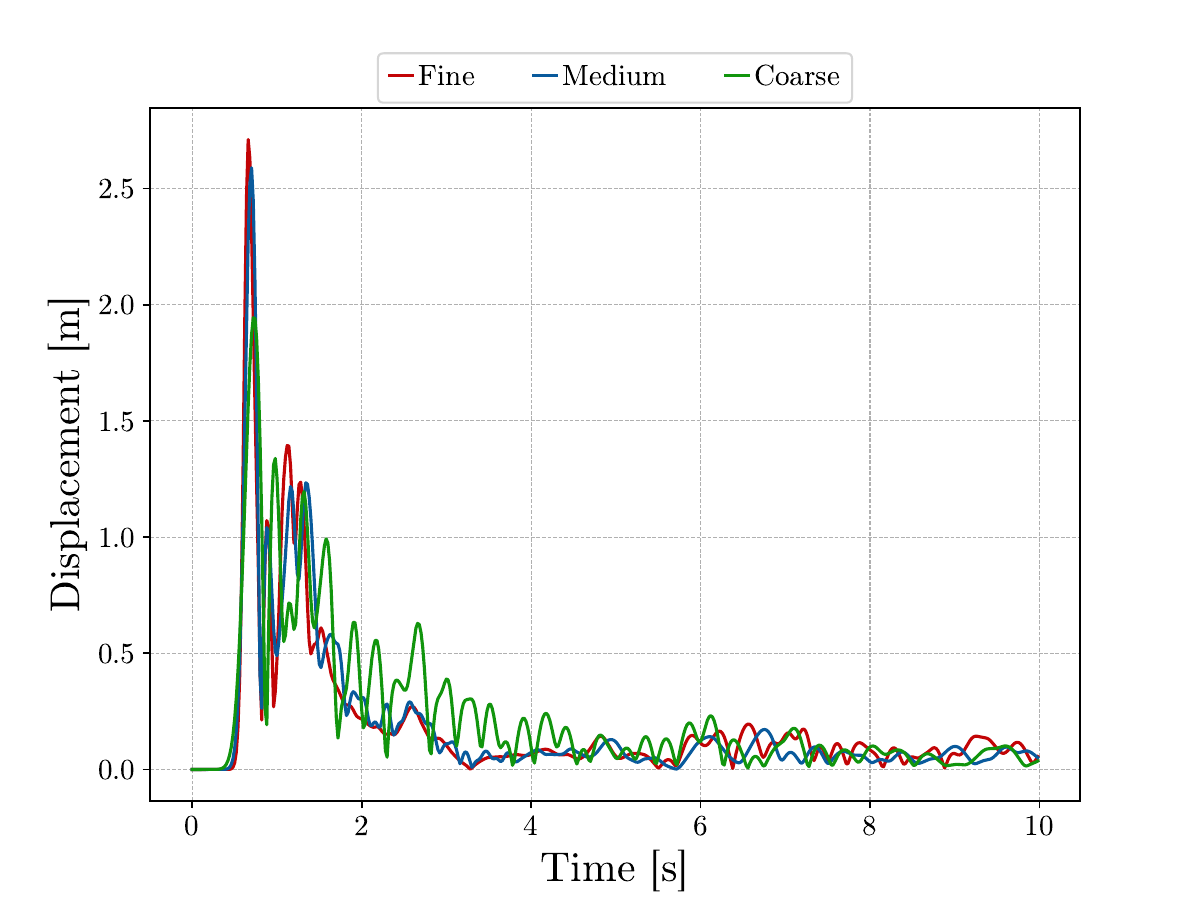}
        \caption{2D elastic soil under point loading -- Test 2: comparisons of displacement magnitudes at receiving point A1 between different mesh sizes.}
         \label{fig:mesh_size_time}
\end{figure}

\revision{Fig.~\ref{fig:mesh_size_time} plots the displacement at receiving point B1 (1125,575) for the three cases. Since particles with exactly the same coordinates do not exist across different mesh sizes, the displacement of the nearest particle was used. From the figure, the coarse case exhibits a considerably smaller peak displacement amplitude compared to the other two cases, which is mostly attributed to the over-damping effects resulting from the larger mesh size. The results for the medium and fine cases show similar trends, with a slight increase in peak magnitude as the mesh is refined. In the coarse case, despite having the smallest peak displacement compared to the other cases, the reflection waves are recorded to be the highest, as shown by Fig.~\ref{fig:mesh_size_time} at around $t=2\sim4$ s. During the time interval of $t = 5\sim10$ s, we qualitatively observe an improvement as the mesh is refined. However, it is evident that even in the finest mesh case, the refraction wave persists. Unfortunately, the error reduction with mesh refinement does not follow the desired convergence trend. As discussed earlier, this is likely due to inaccuracies in dampening surface waves and evanescent waves, which have not yet been addressed in our proposed implementation.}


\subsection{Varying the length of PML domain, $L$}
\label{app:changing_L}

\begin{figure}[h!]
    \centering
     \begin{subfigure}[b]{0.45\textwidth}
         \centering
         \includegraphics[width=\textwidth]{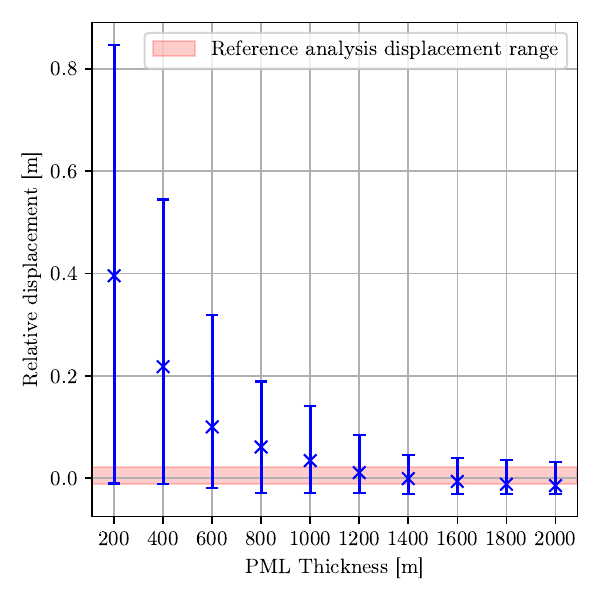}
         \caption{}
         \label{fig:varying_L}
     \end{subfigure}
     \hspace{0.1cm}
     \begin{subfigure}[b]{0.45\textwidth}
         \centering
        \includegraphics[width=\linewidth]{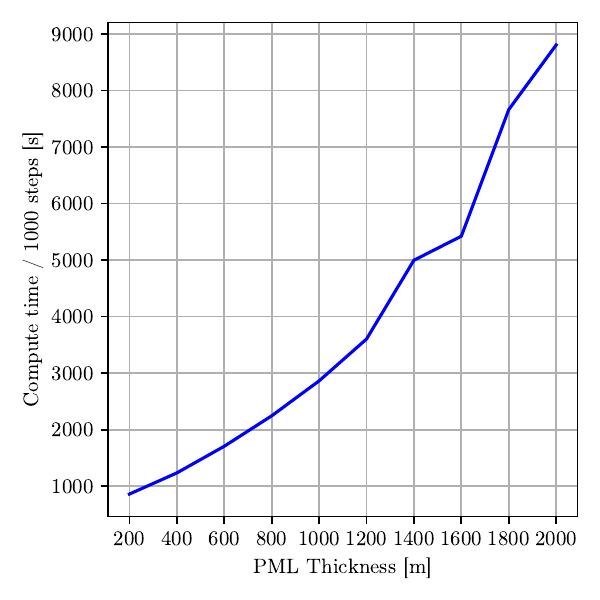}
        \caption{}
        \label{fig:spent_time}
     \end{subfigure}
    \caption{2D elastic soil under point loading -- Test 2: (a) difference of measured displacement magnitudes between $t=5\sim 10$ s with the reference solution for different sizes of $L$, (b) average time spent per 1000 simulation steps.}
    \label{fig:PML_length_all}
\end{figure}

\revision{In this section, we examined the impact of the PML domain size, $L$, on the damping performance of the proposed method. This is specifically related to the number of PML elements over $L$ as we keep the mesh size constant. In the analysis conducted in \Cref{elastic_2d}, the thickness $L$ of the PML domain was set to 1000 m, corresponding to 50 PML elements. Here, we performed additional analyses by varying $L$ from 200 $\sim$ 2000 m. Fig.~\ref{fig:varying_L} presents the recorded displacement at receiving point A1 (2525,65) relative to the reference simulation. For each analysis of $L$, we measured the maximum, minimum, and time-averaged displacement magnitudes over the interval $t = 5\sim10$ s. These values are then subtracted from the time-averaged displacement of the reference case for the same time interval. Since the reference analysis does not yield exactly zero displacement, primarily due to discretization errors inherent in the MPM, the displacement range of the reference analysis is highlighted with a red band to provide an estimate of the maximum accuracy.}

\revision{Fig.~\ref{fig:varying_L} demonstrates that as the PML domain size increases, the amplitude of the reflection waves decreases. Additionally, for  $L \geq 1400$ m, there is no significant difference in the results, indicating convergence. Furthermore, the reduction of errors begins to plateau at $L \approx 800$ m. On the other hand, Fig.~\ref{fig:spent_time} illustrates the average computation time required to complete 1000 steps for each case, revealing an expected increasing trend in analysis time with the increase of $L$. The environment used to conduct the measurement is a PC with an Intel Core i9-13900KF Processor (@5.80 GHz). This suggests that larger PML domains incur higher computational costs without proportional gains in damping effectiveness. Therefore, considering the balance between computation time and damping performance, we selected a PML element number of 40 for the analyses in Sections \ref{elasticplastic_slope} and \ref{slope_failure}.}

\subsection{Varying PML damping parameters, $\alpha_j$ and $\beta$}

\begin{figure}[h!]
    \centering
     \begin{subfigure}[b]{0.45\textwidth}
         \centering
         \includegraphics[width=\textwidth]{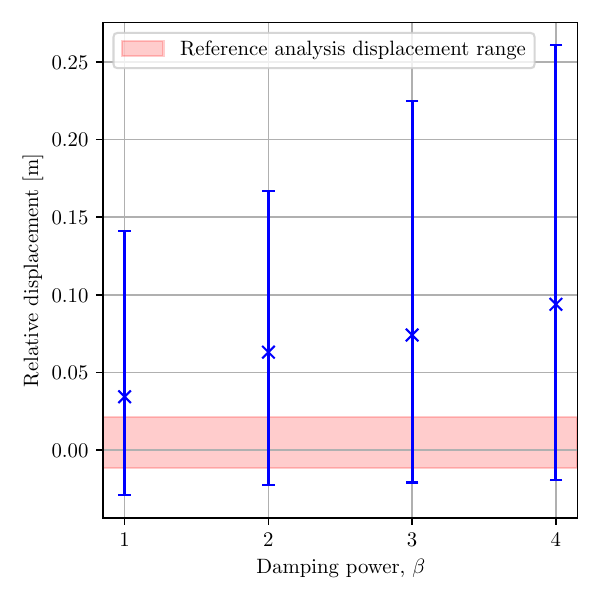}
         \caption{}
         \label{fig:varying_beta}
     \end{subfigure}
     \hspace{0.1cm}
     \begin{subfigure}[b]{0.45\textwidth}
         \centering
        \includegraphics[width=\linewidth]{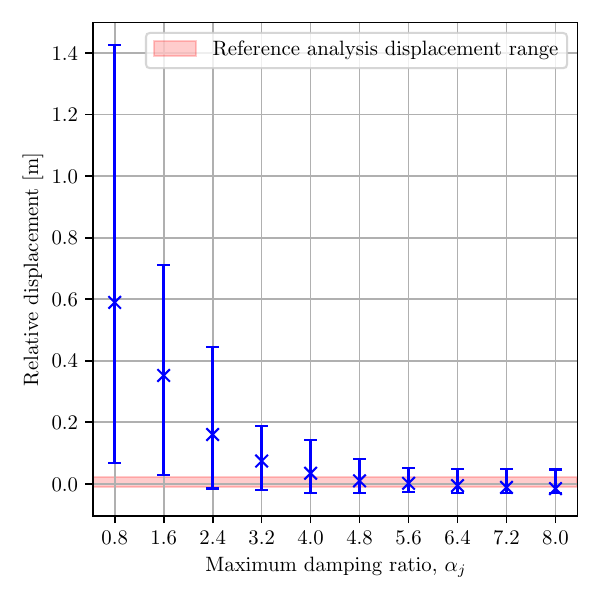}
        \caption{}
        \label{fig:varying_alpha}
     \end{subfigure}
    \caption{2D elastic soil under point loading -- Test 2: difference of measured displacement magnitudes between $t=5\sim 10$ s with the reference solution for different values of (a) $\beta$ and (b) $\alpha_j$.}
    \label{fig:varying_pml_parameters}
\end{figure}

\revision{Next, we examine the PML parameters, specifically the damping power, $\beta$, and the maximum damping ratio, $\alpha_j$. First, we conducted analyses by varying $\beta$ from 1 to 4. The results are shown in Fig.~\ref{fig:varying_beta} in a similar fashion to those presented earlier in Fig.~\ref{fig:varying_L}. The plot shows that although the trend is not significant, the best results are obtained when $\beta$ is set to 1. Mathematically, this is because the integrated damping capacity over the PML length $L$ is maximized when the polynomial $\beta$ is set to linear. We then performed analyses by varying the maximum damping ratio, $\alpha_j$. In this case, we kept the ratio $\alpha_j/L$ constant at 0.004 to avoid abrupt changes in the damping ratio, which tend to induce reflection waves. The analyses were carried out with $\alpha_j$ values of 0.8, 1.6, 2.4, 3.2, 4.0, 4.8, 5.6, 6.4, 7.2, and 8.0 (with $L = 200$, 400, 600, 800, 1000, 1200, 1400, 1600, 1800, and 2000 m, respectively). The results are presented in Fig.~\ref{fig:varying_alpha}. It is observed that the trend is similar to the results obtained by simply changing $L$ as presented in Fig.~\ref{fig:varying_L}. However, up to $\alpha_j = 2.4$, the amplitudes are measured to be larger relative to Fig.~\ref{fig:varying_L}, which is performed with $\alpha_j=4$. On the other hand, we also observe a non-significant change in accuracy for $\alpha_j \geq 3.2$. Based on these results, $\alpha_j = 4$ and $\beta=1$ are generally recommended as the optimal PML parameters. These findings suggest the same values as the one proposed previously by \citet{chen2022}.}

\subsection{Varying visco-elastic and Rayleigh damping parameters}

\revision{We further examined the effects of Rayleigh damping and the viscoelastic model. \citet{chen2022} introduced two overdamping terms to achieve additional damping effects at the PML elements. Here, we investigate the differences in results by varying the relaxed Young’s modulus $E_0$ and Rayleigh damping $\alpha_M$. First, the analysis in \Cref{elastic_2d} assumed $E_0 = 0.99 E$ and $\alpha_M=1.0$ by default, considering the value suggested by \citet{chen2022}. We then performed sensitivity analyses by setting $E_0$ = 0.999$E$ and 0.9$E$. Additionally, we varied the Rayleigh damping factor by setting $\alpha_M$ to 0.1, 0.5, and 0.8. In total, we carried out nine simulation cases, which are summarized in \Cref{tab:simulation_cases}.}

\begin{table}[htbp]
    \centering
    \caption{2D elastic soil under point loading -- Test 2: summary of simulation cases involving visco-elastic model and Rayleigh damping.}
    \label{tab:simulation_cases}
    \begin{tabular}{||l|c||}
        \hline
        \textbf{Case} & \textbf{Description} \\ 
        \hline \hline
        Case 1 & Without Viscoelastic Model, $\alpha_M=1.0$ \\
        Case 2 & \( E_0 = 0.999E \), $\alpha_M=1.0$ \\
        Case 3 & \( E_0 = 0.99E \), $\alpha_M=1.0$ (\textbf{Default}) \\
        Case 4 & \( E_0 = 0.9E \), $\alpha_M=1.0$ \\
        Case 5 & Without Rayleigh Damping, $E_0 = 0.99E$ \\
        Case 6 & \( \alpha_M = 0.1 \), $E_0 = 0.99E$ \\
        Case 7 & \( \alpha_M = 0.5 \), $E_0 = 0.99E$ \\
        Case 8 & \( \alpha_M = 0.8 \), $E_0 = 0.99E$ \\
        Case 9 & Without Rayleigh Damping and Viscoelastic Model \\
        \hline
    \end{tabular}
\end{table}

\begin{figure}[h!]
    \centering
    \includegraphics[width=0.5\linewidth]{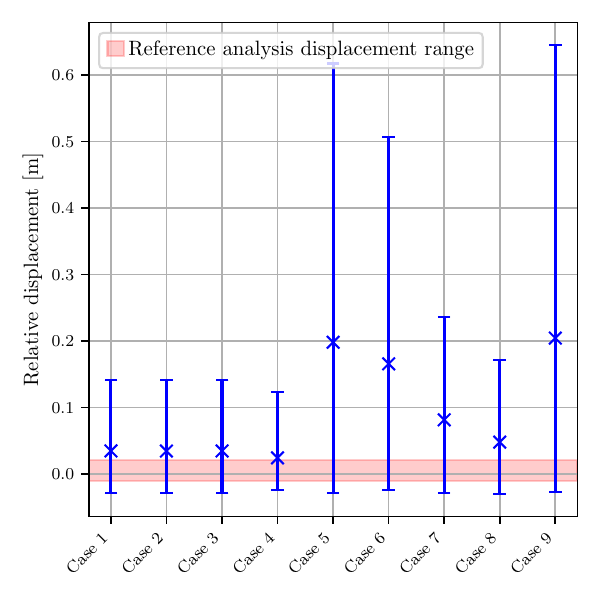}
    \caption{2D elastic soil under point loading -- Test 2: the effects of Rayleigh damping and the viscoelastic model parameters.}
    \label{fig:varying_add_damping}
\end{figure}

\revision{The results compared to the reference solution are shown in Fig.~\ref{fig:varying_add_damping}. For Cases 1–4, Figure~\ref{fig:varying_add_damping} demonstrates that when the difference in magnitude between Young’s modulus and the relaxed Young’s modulus is small, the simulation results show no significant variation. When $E_0 = 0.9E$, the relative displacement slightly decreases due to higher viscoelastic damping. In contrast, removing or reducing the Rayleigh damping factor $\alpha_M$ leads to a significant increase in the amplitude and standard deviation of reflected waves. Conversely, increasing the Rayleigh damping ratio results in a noticeable reduction in the measured relative displacement. In general, setting $E_0$ to a much lower value or $\alpha_M$ to a much higher value is not recommended, as excessive damping in the PML domain can introduce refraction waves from the PML interface into the main analytical domain, potentially causing undesired dynamic vibrations in the system.}

\bibliography{mybibfile}

\end{document}